\documentclass[10pt]{article}
\usepackage[cp1251]{inputenc}
\usepackage[dvips]{epsfig}   \usepackage[dvips]{changebar}
\usepackage[dvips]{graphicx}

\def\lsim{\
  \lower-1.2pt\vbox{\hbox{\rlap{$<$}\lower5pt\vbox{\hbox{$\sim$}}}}\ }
\def\gsim{\
  \lower-1.2pt\vbox{\hbox{\rlap{$>$}\lower5pt\vbox{\hbox{$\sim$}}}}\ }
   
  \usepackage{amsmath,enumerate, amsfonts, amssymb,amsthm}

\begin{document}
\title{ Nature of Lieb's  ``hole'' excitations  \\ and two-phonon states of a Bose gas}
 \author{Maksim Tomchenko
\bigskip \\ {\small Bogolyubov Institute for Theoretical Physics} \\
 {\small 14b, Metrolohichna Str., Kyiv 03143, Ukraine} \\
  {\small E-mail:mtomchenko@bitp.kiev.ua}}
 \date{\empty}
 \maketitle
 \large
 \sloppy
\textit{It is generally accepted that the  ``hole'' and ``particle''
excitations are two independent types of excitations of a
one-dimensional system of point bosons. We  show for a weak coupling
that the Lieb's ``hole'' with the momentum $p=j2\pi/L$ is $j$
identical interacting phonons with the momentum $2\pi/L$ (here, $L$
is the size of the system, and $\hbar=1$). We prove this assertion
for $j=1, 2$ by comparing solutions for a system of point bosons
with solutions for a system of nonpoint bosons obtained in the limit
of the point interaction. The additional arguments show that our
conclusion should be true for any $j=1, 2,  \ldots, N$. Thus, at a
weak coupling, the holes are not a physically independent type of
quasiparticles. Moreover, we find the solution for two interacting
phonons in a Bose system with an interatomic potential of the
general form at a weak coupling and any dimension (1, 2, or 3). It
is also shown for a weak coupling that the largest number of phonons
in a Bose system is equal to the number of atoms $N$. Finally, we
have studied the structure of wave functions for the
Tonks--Girardeau gas and found that the properties
of quasiparticles in this regime are quite strange. } 

 \section{Introduction}
This work is devoted to two main problems: the determination of the
wave function and the energy of two interacting phonons in a Bose
gas with a potential of the general form and the study of the nature
of Lieb's ``holes''. The first problem was not solved, to our
knowledge, and can help one to solve the second problem.

The elementary excitations of a one-dimensional (1D) system of point
bosons are usually separated into two types: particle-like
(``particles'') and hole-like (``holes'')
\cite{lieb1963,lieb1965,yangs1969,gaudinm,takahashi1999,bloch2008,cazalilla2011,astra2013,pustilnik2014,risti2014,karpiuk2015,guan2015}.
At the weak coupling, the dispersion law of ``particles'' coincides
with the Bogolyubov law \cite{bog1947,bz1955} and agrees with
Feynman's solutions \cite{fey1954,fc1956,fey1972} and the more later
models
\cite{brueck1959,cbf-jf,cbf-lee2,campbell1976,yuv2,swf1994a,swf1994b,mt2006,krot2015}
(other references can be found in reviews \cite{reatto,revAJP})
allowing one to describe the microscopic properties of a Bose system
at the weak and intermediate couplings. Therefore, it is natural to
consider that the particles correspond to Bogolyubov--Feynman
quasiparticles. The dispersion law of holes was found only in the
approach based on the Bethe ansatz, in the well-known work by Lieb
\cite{lieb1963}. In this case, Lieb attacked the Bogolyubov's and
Feynman's approaches and proposed some arguments in favor of that
the holes are a physically independent type of elementary
excitations \cite{lieb1963,lieb1965}. This point of view became
traditional. Later on, it was found that the dispersion law of holes
is close to that for the soliton solution of the 1D
Gross--Pitaevskii equation \cite{tsuzuki1971,ishikawa1980}. This
became the main argument in favor of that the holes are a particular
independent type of quasiparticles. However, such point of view does
not agree with the models
\cite{bog1947,bz1955,fey1954,fc1956,fey1972,brueck1959,cbf-jf,cbf-lee2,campbell1976,yuv2,swf1994a,swf1994b,mt2006,krot2015}.
It is  important that the models
\cite{bz1955,fey1954,fc1956,campbell1976,yuv2,swf1994a,swf1994b,mt2006,krot2015}
\textit{work} in 1D, since they do not use a condensate (we note
that the Bogolyubov's method also works in 1D at small $\gamma$ and
$T$, if $N$ is finite \cite{mtujp2019}). If the holes would be a
separate type of quasiparticles, this would mean the significant
shortcoming in the models
\cite{bog1947,bz1955,fey1954,fc1956,fey1972,brueck1959,cbf-jf,cbf-lee2,campbell1976,yuv2,swf1994a,swf1994b,mt2006,krot2015}
and in close ones. In addition, if the holes are an independent type
of excitations, then they must give a separate contribution to
thermodynamic quantities (since holes interact with particles and,
therefore, participate in the thermal equilibrium). Such analysis
indicates that the question about the nature of holes is very
important.

The one-dimensional system differs qualitatively from a
three-dimensional (3D) one by that the atom in a 1D system cannot
get around another atom. The former can only pass \textit{through}
the latter. Despite this circumstance, Lieb believed that 1D and 3D
systems are qualitatively similar \cite{lieb1963}. Therefore, he
made conclusion \cite{lieb1963} that holes can exists also in 3D
systems, at least in case of a strong coupling.

In what follows, we will study the structure of the wave functions
of ``particles'' and holes and will show that, at a weak coupling,
the hole is a collection of interacting ``particles''.  It was noted
in the literature that the holes are not a mathematically
independent type of excitations
\cite{lieb1963,cazalilla2011,mtjltp2017}. This conclusion was based
on the Lieb--Liniger equations. However, these equations are not
enough to clarify the physical nature of holes.

Let us consider what the Lieb--Liniger equations can say about the
nature of holes. These equations describe a periodic 1D system of
spinless point bosons \cite{ll1963}. Gaudin wrote them in the form
\cite{gaudinm,gaudin1971}
\begin{eqnarray}
Lk_{i}= 2\pi
n_{i}+2\sum\limits_{j=1}^{N}\arctan{\frac{c}{k_{i}-k_{j}}}|_{j\neq
i}, \ i=1,\ldots, N,
   \label{1} \end{eqnarray}
where $N$ is the number of bosons, $L$ is the size of the system,
and $n_{i}=0, \pm 1, \pm 2, \ldots$. In the literature, the point
bosons are usually described by the Lieb--Liniger equations in the
Yang--Yang's form \cite{yangs1969}:
\begin{eqnarray}
Lk_{i}=2\pi
I_{i}-2\sum\limits_{j=1}^{N}\arctan{\frac{k_{i}-k_{j}}{c}}, \quad
i=1,\ldots, N. \label{2} \end{eqnarray} The  equations (\ref{1}) and
(\ref{2}) are equivalent \cite{gaudinm,gaudin1971}: the formula
$\arctan{\alpha}=(\pi/2)sgn(\alpha)-\arctan{(1/\alpha)}$ allows one
to rewrite Eqs. (\ref{2}) in the form (\ref{1}). In this case,
\begin{equation}
I_{i}=n_{i}+i-\frac{N+1}{2}.
     \label{In} \end{equation}
The ground state of the system corresponds to the quantum numbers
$\{I_{i}\}=(1-\frac{N+1}{2},2-\frac{N+1}{2},\ldots,N-\frac{N+1}{2})$,
the particle-like excitation with the momentum $p=2\pi j/L$
corresponds to
$\{I_{i}\}=(1-\frac{N+1}{2},\ldots,N-1-\frac{N+1}{2},N-\frac{N+1}{2}+j)$,
and a hole with the momentum $p=2\pi l/L$ ($l>0$) corresponds to the
quantum numbers $I_{i\leq N-l}=i-\frac{N+1}{2}$, $I_{i>
N-l}=1+i-\frac{N+1}{2}$. In the language of Eqs. (\ref{1}), those
states correspond to the following collections of quantum numbers
$\{n_{i}\}=(n_{1},\ldots,n_{N})$: $(0,\ldots,0)$, $(0,\ldots,0,j)$,
and $(0,\ldots,0,1,\ldots,1)$, where $1$ is repeated  $l$ times. In
this case, the state $(0,\ldots,0,1)$ is particular: it can be
considered as a particle and as a hole. In the last case, any state
$(n_{1},\ldots,n_{N})$ can be considered as a collection of
interacting holes. If the state $(0,\ldots,0,1)$ is a particle, then
any state   can be considered as a collection of interacting
particles. Therefore, the physical nature of the state
$(0,\ldots,0,1)$ is the \textit{key point}. From physical
reasonings, we may expect that the state $(0,\ldots,0,1)$
corresponds to a phonon with the wavelength $\lambda=L$ (indeed, if
the state $(0,\ldots,0,1)$ would correspond to a hole, then the
phonon with $\lambda=L$ would be absent in the system, which is
strange). In this case, each state $(n_{1},\ldots,n_{N})$ can be
considered as a collection of interacting phonons. In particular,
the state $(0,\ldots,0,j)$ should correspond to one phonon with the
momentum $p=2\pi j/L$. As for the state with $n_{j\leq N-l}=0,
n_{j\geq N-l+1}=1,$ it should correspond to $l$ interacting phonons,
each of them has the wavelength $\lambda=L$ and the momentum
$2\pi/L$. However, according to the Lieb's classification
\cite{lieb1963}, the state with the quantum numbers $n_{j\leq
N-l}=0, n_{j\geq N-l+1}=1$ corresponds to a \textit{hole} with the
momentum $p=2\pi l/L> 0$. Therefore, such hole  should coincide with
$l$ interacting phonons, each of them has the momentum $2\pi/L$.
This possibility is also seen from the analysis by Lieb
\cite{lieb1963}.

To ascertain the nature of a hole, it is necessary to study the
structure of $N$-boson wave functions of a hole and a particle. In
what follows, we will prove for a weak coupling that the state
$(0,\ldots,0,1)$ corresponds to a phonon, and the hole with the
momentum $p=4\pi/L$ coincide with two interacting phonons
$(0,\ldots,0,1)$. We will also show that, at a strong coupling, the
structure of quasiparticles is more unusual.

\section{Phonon with the quantum numbers $\{n_{i}\}=(0,\ldots,0,1)$.}
One can investigate the structure of wave functions of a
``particle'' and a hole in two ways: based on the wave functions of
point bosons \cite{gaudinm,ll1963} or on the wave functions of
nonpoint bosons (i.e., bosons with nonzero interaction radius)
\cite{bz1955,fey1954,fc1956,yuv2,krot2015,jastrow,woo,feenberg,yuv1},
by passing to a point potential in the last case. Let us consider
the second way.

Consider a periodic system of $N$ bosons with interatomic potential
of the general form $U(\textbf{r}_{j}-\textbf{r}_{l})$. The
dimensionality  can be equal to $1$, $2$, or $3$.  The ground state
of a gas is described by the wave function \cite{yuv1}
\begin{eqnarray}
   \Psi_{0}(\textbf{r}_1,\ldots ,\textbf{r}_N) =A_{0} e^{S(\textbf{r}_1,\ldots ,\textbf{r}_N)},
    \label{2-1}   \end{eqnarray}
\begin{eqnarray}
  S&=&
   \sum\limits_{\textbf{q}_{1}\neq 0}\frac{a_{2}(\textbf{q}_{1})}{2!}\rho_{\textbf{q}_{1}}
   \rho_{-\textbf{q}_{1}}+ \sum\limits_{\textbf{q}_{1},\textbf{q}_{2}\neq 0}^{\textbf{q}_{1}+\textbf{q}_{2}\not= 0}
  \frac{a_{3}(\textbf{q}_{1},\textbf{q}_{2})}{3!N^{1/2}}
 \rho_{\textbf{q}_{1}}\rho_{\textbf{q}_{2}}\rho_{-\textbf{q}_{1} - \textbf{q}_{2}}+\ldots +
   \nonumber\\  &+&
  \sum\limits_{\textbf{q}_{1},\ldots,\textbf{q}_{N-1}\neq 0}^{\textbf{q}_{1}+\ldots +\textbf{q}_{N-1}\not= 0}
  \frac{a_{N}(\textbf{q}_{1},\ldots,\textbf{q}_{N-1})}{N!N^{(N-2)/2}}
 \rho_{\textbf{q}_1}\ldots\rho_{\textbf{q}_{N-1}}
 \rho_{-\textbf{q}_{1} - \ldots - \textbf{q}_{N-1}},
    \label{2-2}   \end{eqnarray}
and the wave function of a one-phonon state reads \cite{yuv2}
 \begin{equation}
    \Psi_{\textbf{p}}(\textbf{r}_1,\ldots ,\textbf{r}_N) =
  A_{\textbf{p}}\psi_{\textbf{p}}\Psi_0,
  \label{2-3}     \end{equation}
       \begin{eqnarray}
 \psi_{\textbf{p}} & =&
  b_{1}(\textbf{p})\rho_{-\textbf{p}} +
 \sum\limits_{\textbf{q}_{1}\neq 0}^{\textbf{q}_{1}+\textbf{p}\neq 0}
  \frac{b_{2}(\textbf{q}_{1};\textbf{p})}{2!N^{1/2}}
 \rho_{\textbf{q}_{1}}\rho_{-\textbf{q}_{1}-\textbf{p}} +  \sum\limits_{\textbf{q}_{1},\textbf{q}_{2}\neq 0}^{\textbf{q}_{1}+
 \textbf{q}_{2}+\textbf{p} \not= 0}
  \frac{b_{3}(\textbf{q}_{1},\textbf{q}_{2};\textbf{p})}{3!N}
 \rho_{\textbf{q}_{1}}\rho_{\textbf{q}_{2}}\rho_{-\textbf{q}_{1}-\textbf{q}_{2}-\textbf{p}}
 + \nonumber \\ &+& \ldots + \sum\limits_{\textbf{q}_{1},\ldots,\textbf{q}_{N-1}\neq 0}^{\textbf{q}_{1}+\ldots +\textbf{q}_{N-1}+\textbf{p}\not= 0}
  \frac{b_{N}(\textbf{q}_{1},\ldots,\textbf{q}_{N-1};\textbf{p})}{N!N^{(N-1)/2}}
 \rho_{\textbf{q}_1}\ldots\rho_{\textbf{q}_{N-1}}
 \rho_{-\textbf{q}_{1} - \ldots - \textbf{q}_{N-1}-\textbf{p}}.
       \label{2-4}\end{eqnarray}
Here, $N$ is the total number of atoms, $\textbf{r}_j$ are the
coordinates of atoms, $A_{0}$ and $A_{\textbf{p}}$ are the
normalization constants, $\rho_{\textbf{q}}$ are the collective
variables
\begin{equation}
   \rho_{\textbf{q}} = \frac{1}{\sqrt{N}}\sum\limits_{j=1}^{N}e^{-i\textbf{q}\textbf{r}_j},
 \label{rok}    \end{equation}
and all wave vectors $\textbf{q}_l$, $\textbf{p}_l$, $\textbf{p}$
are quantized in the 3D case by the rule
 \begin{equation}
  \textbf{q}=2\pi \left (\frac{j_{x}}{L_{x}}, \frac{j_{y}}{L_{y}},
  \frac{j_{z}}{L_{z}} \right ),
    \label{2-6} \end{equation}
where $j_{x}, j_{y}, j_{z}$ are integers, and  $L_{x}, L_{y}, L_{z}$
are the sizes of the system.  Equations (\ref{2-2}), (\ref{2-4}) are
exact.  Note that, in \cite{yuv2,yuv1} series (\ref{2-2}),
(\ref{2-4}) tend to infinity (i.e., the sums with  $a_{j}, b_{j}$
for $j=N+1,\ldots, \infty,$ are taken into account). We will see in
Appendix 1 that these series must break down according to
(\ref{2-2}), (\ref{2-4}).

The approximate solutions for the functions $\Psi_{0}$ and
$\Psi_{\textbf{p}}$ were obtained by Feynman \cite{fey1954,fc1956},
Bogolyubov and Zubarev \cite{bz1955}, and Jastrow \cite{jastrow}.
Then these methods were developed in a lot of works (see
\cite{cbf-jf,cbf-lee2,campbell1976,yuv2,swf1994a,swf1994b,mt2006,krot2015,woo,feenberg,yuv1}
and reviews \cite{reatto,revAJP}). We will base on the collective
variables method by Vakarchuk and Yukhnovskii \cite{yuv2,yuv1}. It
allows one to get two exact chains of equations for the functions
$a_{j}$ and $b_{j}$ at $N=\infty$. The first equations from those
chains are given in Appendix 2.

The wave function (\ref{2-3}), (\ref{2-4}) with small $|p|$ can be
considered as the definition of a phonon (here, $b_{j}\sim N^{0}$
for all $j$). Basic is the zero approximation
$\psi_{\textbf{p}}=b_{1}\rho_{-\textbf{p}}$; the corrections can be
found from the Schr\"{o}dinger equation. Such solution for a phonon
was studied theoretically in many works, starting from
\cite{fey1954,bz1955,fey1972,yuv2,bijl}, and the results agree with
experiments. The properties of collective variables \cite{yuv1}
imply that function (\ref{2-3}), (\ref{2-4}) describes also $l$
interacting phonons with the total momentum
$\textbf{p}=\textbf{p}_{1}+\ldots + \textbf{p}_{l}$, if we make the
following changes in (\ref{2-3}), (\ref{2-4}):
$\Psi_{\textbf{p}}\rightarrow \Psi_{\textbf{p}_{1}\ldots
\textbf{p}_{l}}$, $\psi_{\textbf{p}}\rightarrow
\psi_{\textbf{p}_{1}\ldots \textbf{p}_{l}}$,
$A_{\textbf{p}}\rightarrow A_{\textbf{p}_{1}\ldots \textbf{p}_{l}}$,
$b_{j}(\textbf{q}_{1},\ldots,\textbf{q}_{j-1};\textbf{p})\rightarrow
b_{j}(\textbf{q}_{1},\ldots,\textbf{q}_{j-1};\textbf{p}_{1},\ldots,
\textbf{p}_{l},N)$ for all $j$ (now, $b_{j}$ depend on $N$,
generally speaking). We can verify that, in this case, $-i\hbar
\sum\limits_{j}\frac{\partial}{\partial
\textbf{r}_{j}}\Psi_{\textbf{p}_{1}\ldots \textbf{p}_{l}}= \hbar
\textbf{p}\Psi_{\textbf{p}_{1}\ldots \textbf{p}_{l}}$.

For the weak coupling ($\gamma \ll 1$), we can set $a_{j\geq 3}=0$,
$b_{j\geq 2}=0$ (it is the zero approximation; here, $\gamma=
\rho^{1/3}c m/\hbar^{2}$ (for 3D), $c m/\hbar^{2}$ (2D), $2c
m/(\rho\hbar^{2})$ (1D), $\rho=N/V$ is the  particle number density,
and $c=\nu(0)/2$, see (\ref{2-9})). The coefficient
$b_{1}(\textbf{p})$ is considered to be normalizing: we set
$b_{1}(\textbf{p})=1$. Then the equations in Appendix 2 yield
\cite{yuv2,yuv1}
\begin{equation}
a_{2}(\textbf{p})\equiv a_{2}(p) =\frac{1-\alpha_{\textbf{p}}}{2},
\quad
\alpha_{\textbf{p}}=\sqrt{1+\frac{2\rho\nu(p)}{\hbar^{2}p^{2}/(2m)}},
     \label{2-7} \end{equation}
\begin{equation}
E(\textbf{p}) =\frac{\hbar^{2}p^{2}}{2m}(1-2a_{2}(\textbf{p}))=
\sqrt{\left (\frac{\hbar^{2}p^{2}}{2m}\right )^{2}+2\rho\nu(p)\left
(\frac{\hbar^{2}p^{2}}{2m}\right )}\equiv E_{B}(\textbf{p}),
     \label{2-8} \end{equation}
\begin{equation}
 \nu(\textbf{p}) = \int\limits_{-L_{x}}^{L_{x}}dx \int\limits_{-L_{y}}^{L_{y}}dy
 \int\limits_{-L_{z}}^{L_{z}}dz U(r)e^{-i\textbf{p}\textbf{r}}.
        \label{2-9} \end{equation}
We have obtained the Bogolyubov dispersion law. Formula (\ref{9-1})
from Appendix 2 gives the known Bogolyubov solution for the
ground-state energy $E_{0}$ \cite{bog1947}.

In the zero approximation the sound velocity is
$v_{s}=\sqrt{\frac{\rho\nu(0)}{m}} \equiv v_{s}^{(0)}$. In the next
approximation the solution is as follows \cite{yuv2}:
\begin{equation}
 v_{s}=v_{s}^{(0)}(1 +\delta_{s}), \quad \delta_{s} = -\frac{\hbar^{2}}{32m^{2}(v_{s}^{(0)})^{2}}
 \frac{1}{N}\sum\limits_{\textbf{q}\neq 0}\frac{q^{2}}{\alpha_{\textbf{q}}^{3}}\left (\frac{2\rho\nu(q)}{\hbar^{2}q^{2}/(2m)}\right )^{2}.
        \label{2-10} \end{equation}
For a 1D system the energy of a phonon with the momentum $\hbar
p_{1}=\hbar 2\pi/L$ is $E(p_{1})=\hbar p_{1}v_{s}$. In this case,
for a \textit{finite} system we should set
$v_{s}^{(0)}=\sqrt{\frac{\rho\nu(p_{1})}{m}+\frac{\hbar^{2}p_{1}^{2}}{4m^{2}}}$.

Consider a finite 1D system of point bosons ($U(r)=2c\delta(r)$,
$\nu(p)=2c$) and set $\hbar=2m=1$, $\gamma =c/\rho$. The
above-presented formulae give the energy of a phonon with the
momentum $p_{1}=2\pi/L$:
\begin{equation}
E(p_{1})=\sqrt{p_{1}^{4}+4\rho^{2}\gamma p_{1}^{2}}\cdot (1
+\delta_{s})=\frac{4\pi
\rho\sqrt{\gamma}}{L}\sqrt{1+\frac{\pi^{2}}{\gamma N^{2}}}\cdot (1
+\delta_{s}),
        \label{2-11} \end{equation}
\begin{equation}
\delta_{s} = -\frac{1}{4N}\frac{1}{1+\pi^{2}/(\gamma N^{2})}
 \sum\limits_{j=1,2,\ldots,\infty}\frac{1}{1+\pi^{2}j^{2}/(\gamma N^{2})}\frac{1}{\sqrt{1+\gamma N^{2}/(\pi^{2}j^{2})}}.
        \label{2-12} \end{equation}
These formulae are valid for $N^{-2} \ll \gamma \ll 1$.

Our task is to clarify the nature of the particle $(0,\ldots,0,1)$.
It is known that the energy $E_{L}(p)$ of Lieb's particle for small
$p$ is close to the Bogolyubov energy $E_{B}(p)$ (\ref{2-8}). The
small deviation of the particle energy from $E_{B}(p)$ contains the
information about the nature of the particle. Let us represent the
energy of the particle with the momentum $p_{1}= 2\pi/L$ in the form
(\ref{2-11}):
\begin{equation}
E_{L}(p_{1})=\frac{4\pi
\rho\sqrt{\gamma}}{L}\sqrt{1+\frac{\pi^{2}}{\gamma N^{2}}}\cdot (1
+\delta_{sL}).
        \label{2-13} \end{equation}
The energy and momentum of the particle is given by the known
formulae
\begin{equation}
E_{L}(p)=\sum\limits_{i=1}^{N}(\acute{k}^{2}_{i}-k^{2}_{i}),
     \label{2-14} \end{equation}
\begin{equation}
p=\sum\limits_{i=1}^{N}(\acute{k}_{i}-k_{i})=\frac{2\pi}{L}\sum\limits_{i=1}^{N}(\acute{n}_{i}-n_{i}).
     \label{2-15} \end{equation}
In our case, the collections $\{\acute{k}_{i}\}$ and  $\{k_{i}\}$
are solutions of the Gaudin's equations (\ref{1}) for a state with
one particle ($\{\acute{n}_{i}\}=(0,\ldots,0,1)$) and for the ground
state ($\{n_{i}\}=(0,\ldots,0,0)$), respectively. The quasimomenta
$\{\acute{k}_{i}\}$ and $\{k_{i}\}$ can be obtained numerically from
Eqs. (\ref{1}) by the Newton method (the Yang--Yang's equations
(\ref{2}) give the same solution).

\begin{figure}[ht]
\centerline{\includegraphics[width=85mm]{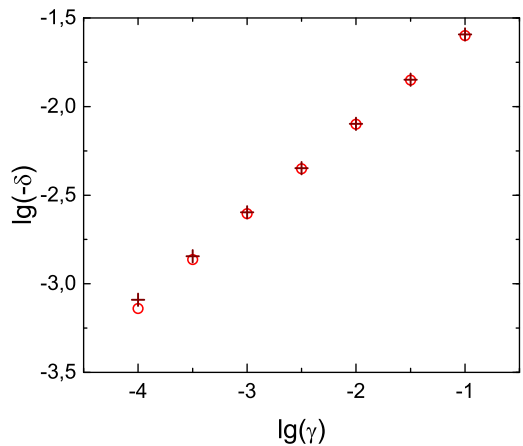} } \caption{
[Color online] Functions $\delta_{s}(\gamma)$ (circles) and
$\delta_{sL}(\gamma)$ (crosses) obtained from Eqs. (\ref{2-12}) and
(\ref{1}), (\ref{2-13})--(\ref{2-15}), respectively; $\rho=1$,
$N=1000$.
 \label{fig1}}
\end{figure}

It is seen from Fig. 1 that the small quantity $\delta_{sL}$
obtained from Eqs. (\ref{2-13})-(\ref{2-15}), (\ref{1}) coincides
with high accuracy with $\delta_{s}$ (\ref{2-12}). The difference of
$\delta_{sL}$ and $\delta_{s}$ is about $1\%$ for
$\gamma=0.0001$--$0.1$. Since the function
$\psi_{\textbf{p}}=\rho_{-\textbf{p}}$ for small $p$ describes a
phonon, we conclude that Lieb's particle $\{n_{i}\}=(0,\ldots,0,1)$
is a phonon. In this case, the Gaudin's equations (\ref{1}) imply
that the hole with the momentum $p=2\pi l/L$ ($l>1$) should coincide
with $l$ interacting phonons with the momentum $2\pi/L$. Let us
verify this directly for $l=2$.

\section{Two interacting phonons vs a hole with the quantum numbers $\{n_{i}\}=(0,\ldots,0,1,1)$.}
In the language of the Lieb--Liniger equations (\ref{2}), the hole
with the momentum $p=4\pi/L$ is characterized by the quantum numbers
$\{I_{i}\}=(-\frac{N-1}{2},-\frac{N-3}{2},\ldots,\frac{N-5}{2},1+\frac{N-3}{2},1+\frac{N-1}{2})$.
In the language of the Lieb--Liniger equations in the Gaudin's form
(\ref{1}), such hole is described by the quantum numbers
$\{n_{i}\}=(0,\ldots,0,1,1)$. In the previous section we proved that
the state $\{n_{i}\}=(0,\ldots,0,1)$ describes a \textit{phonon}
with the momentum $p=2\pi/L$. The state $\{n_{i}\}=(0,\ldots,0,1,0)$
is equivalent to $\{n_{i}\}=(0,\ldots,0,1)$. Therefore, it is
obvious that the state $\{n_{i}\}=(0,\ldots,0,1,1)$ is two
interacting phonons with the momentum $p=2\pi/L$. We now verify this
assumption independently, by using the collective variables method.

Consider a Bose gas with weak coupling and dimensionality of $1, 2,$
or $3$. Let us find the wave function and the energy of two
interacting phonons with wave vectors $\textbf{p}_{1}$ and
$\textbf{p}_{2}$. Feynman noticed that the energy of interaction
($\delta E$) of two phonons should be by $\sim N$ times less than
the energy of one phonon \cite{fey1954}. However, the solutions for
a wave function and $\delta E$ were not found.

The ground state is described by the wave function (\ref{2-1}),
(\ref{2-2}) satisfying the Schr\"{o}dinger equation
\begin{equation}
-\frac{\hbar^{2}}{2m}\sum\limits_{j}\triangle_{j}\Psi +
\frac{1}{2}\sum\limits_{ij}^{i \not= j}
 U(|\textbf{r}_{i}-\textbf{r}_{j}|)\Psi=E\Psi
     \label{3-1} \end{equation}
with energy $E=E_{0}$. The equations for $E_{0}$ and the functions
$a_{j}$ from (\ref{2-2}) are given in Appendix 2.  If the system
contains one phonon, then the wave function is
$\psi_{\textbf{p}}\Psi_0$  with  $\psi_{\textbf{p}}$ (\ref{2-4}),
and the solutions for the functions $b_{j}$ and the energy of a
quasiparticle are given in the previous section. If two phonons with
wave vectors $\textbf{p}_{1}$ and $\textbf{p}_{2}$ are present, then
the system is described by the wave function
$\psi_{\textbf{p}_{1}\textbf{p}_{2}}\Psi_0$. We substitute this
function in the Schr\"{o}dinger equation and take into account that
$\Psi_{0} =Ae^{S}$ satisfies this equation with energy $E_{0}$. As a
result, we obtain the equation for the function
$\psi_{\textbf{p}_{1}\textbf{p}_{2}}$:
\begin{equation}
-\frac{\hbar^{2}}{2m}\sum\limits_{j}\left [
\triangle_{j}\psi_{\textbf{p}_{1}\textbf{p}_{2}} +
2(\nabla_{j}S)(\nabla_{j}\psi_{\textbf{p}_{1}\textbf{p}_{2}})\right
]=E_{\textbf{p}_{1}\textbf{p}_{2}}\psi_{\textbf{p}_{1}\textbf{p}_{2}},
     \label{3-2} \end{equation}
where $E_{\textbf{p}_{1}\textbf{p}_{2}}=E-E_{0}$ is the energy of
two interacting phonons. Since the interaction of two phonons should
be weak, we seek $\psi_{\textbf{p}_{1}\textbf{p}_{2}}$ in the form
\begin{equation}
\psi_{\textbf{p}_{1}\textbf{p}_{2}}=
\psi_{\textbf{p}_{1}}\psi_{\textbf{p}_{2}}+\frac{\delta\psi_{\textbf{p}_{1}\textbf{p}_{2}}}{\sqrt{N}},
     \label{3-3} \end{equation}
where $\psi_{\textbf{p}_{1}}$ and $\psi_{\textbf{p}_{2}}$ are
one-phonon solutions. We substitute
$\psi_{\textbf{p}_{1}\textbf{p}_{2}}$ (\ref{3-3}) in Eq. (\ref{3-2})
and take into account that the one-phonon functions
$\psi_{\textbf{p}_{1}}$ and $\psi_{\textbf{p}_{2}}$ satisfy Eq.
(\ref{3-2}) with the energies $E(\textbf{p}_{1})$ and
$E(\textbf{p}_{2}),$ respectively. In this way we get the following
equation for $\delta\psi_{\textbf{p}_{1}\textbf{p}_{2}}$:
\begin{eqnarray}
&&-\frac{\hbar^{2}}{2m}\sum\limits_{j}\left
[2(\nabla_{j}\psi_{\textbf{p}_{1}})(\nabla_{j}\psi_{\textbf{p}_{2}})
+ \triangle_{j}\delta\psi_{\textbf{p}_{1}\textbf{p}_{2}} +
2(\nabla_{j}S)(\nabla_{j}\delta\psi_{\textbf{p}_{1}\textbf{p}_{2}})\right
]=\nonumber \\ &&=[E(\textbf{p}_{1})+E(\textbf{p}_{2})+\delta
E]\delta\psi_{\textbf{p}_{1}\textbf{p}_{2}}+\delta E
\psi_{\textbf{p}_{1}}\psi_{\textbf{p}_{2}},
     \label{3-5} \end{eqnarray}
\begin{equation}
E_{\textbf{p}_{1}\textbf{p}_{2}}=E(\textbf{p}_{1})+E(\textbf{p}_{2})+\delta
E.
     \label{3-6} \end{equation}
Here, the energy $E_{\textbf{p}_{1}\textbf{p}_{2}}$ of two
interacting phonons is represented as a sum of the energies
$E(\textbf{p}_{1})$ and $E(\textbf{p}_{2})$ of free phonons and the
correction $\delta E$.

The solution for $\delta\psi_{\textbf{p}_{1}\textbf{p}_{2}}$ should
have the form $\psi_{\textbf{p}}$ (\ref{2-4}) with
$\textbf{p}=\textbf{p}_{1}+\textbf{p}_{2}$, since formula
(\ref{2-4}) describes the state with any number of quasiparticles
possessing the total momentum $\hbar\textbf{p}$:
\begin{eqnarray}
 \delta\psi_{\textbf{p}_{1}\textbf{p}_{2}}  &=&
 B_{1}(\textbf{p}_{1},\textbf{p}_{2})\rho_{-\textbf{p}} +
 \sum\limits_{\textbf{q}\neq 0}^{\textbf{q}+\textbf{p}\neq 0}
  \frac{B_{2}(\textbf{q};\textbf{p}_{1},\textbf{p}_{2})}{2!N^{1/2}}
 \rho_{\textbf{q}}\rho_{-\textbf{q}-\textbf{p}}  \nonumber
 \\&+& \sum\limits_{\textbf{q}_{1},\textbf{q}_{2}\neq
 0}^{\textbf{q}_{1}+
 \textbf{q}_{2}+\textbf{p} \not= 0}
  \frac{B_{3}(\textbf{q}_{1},\textbf{q}_{2};\textbf{p}_{1},\textbf{p}_{2})}{3!N}
 \rho_{\textbf{q}_{1}}\rho_{\textbf{q}_{2}}\rho_{-\textbf{q}_{1}-\textbf{q}_{2}-\textbf{p}}
 + \ldots  \nonumber \\ &+& \sum\limits_{\textbf{q}_{1},\ldots,\textbf{q}_{N-1}\neq 0}^{\textbf{q}_{1}+\ldots +\textbf{q}_{N-1}+\textbf{p}\not= 0}
  \frac{B_{N}(\textbf{q}_{1},\ldots,\textbf{q}_{N-1};\textbf{p}_{1},\textbf{p}_{2})}{N!N^{(N-1)/2}}
 \rho_{\textbf{q}_1}\ldots\rho_{\textbf{q}_{N-1}}
 \rho_{-\textbf{q}_{1} - \ldots - \textbf{q}_{N-1}-\textbf{p}},
       \label{3-4}\end{eqnarray}
where $\textbf{p}=\textbf{p}_{1}+\textbf{p}_{2}$. We substitute
$\delta\psi_{\textbf{p}_{1}\textbf{p}_{2}}$ (\ref{3-4}) in
(\ref{3-5}). The result is reduced to the form
\begin{eqnarray}
0  &=&
 C_{1}(\textbf{p}_{1},\textbf{p}_{2})\rho_{-\textbf{p}} +
 \sum\limits_{\textbf{q}\neq 0}^{\textbf{q}+\textbf{p}\neq 0}
  \frac{C_{2}(\textbf{q};\textbf{p}_{1},\textbf{p}_{2})}{N^{1/2}}
 \rho_{\textbf{q}}\rho_{-\textbf{q}-\textbf{p}} + \ldots + \nonumber \\ &+&
 \sum\limits_{\textbf{q}_{1},\ldots,\textbf{q}_{N-1}\neq 0}^{\textbf{q}_{1}+\ldots +\textbf{q}_{N-1}+\textbf{p}\not= 0}
  \frac{C_{N}(\textbf{q}_{1},\ldots,\textbf{q}_{N-1};\textbf{p}_{1},\textbf{p}_{2})}{N^{(N-1)/2}}
 \rho_{\textbf{q}_1}\ldots\rho_{\textbf{q}_{N-1}}
 \rho_{-\textbf{q}_{1} - \ldots - \textbf{q}_{N-1}-\textbf{p}}
       \label{3-7}\end{eqnarray}
($\textbf{p}=\textbf{p}_{1}+\textbf{p}_{2}$). Since
$\rho_{-\textbf{p}}$,
$\rho_{\textbf{q}}\rho_{-\textbf{q}-\textbf{p}}$,
$\rho_{\textbf{q}_{1}}\rho_{\textbf{q}_{2}}\rho_{-\textbf{q}_{1}-\textbf{q}_{2}-\textbf{p}},
\ldots $ are independent functions of the variables
$\textbf{r}_1,\ldots ,\textbf{r}_N$ \cite{yuv1}, Eq. (\ref{3-7}) is
equivalent to the system of $N$ equations
\begin{eqnarray}
C_{j}(\textbf{q}_{1},\ldots,\textbf{q}_{j-1};\textbf{p}_{1},\textbf{p}_{2})=0,
\quad j=1,\ldots, N.
       \label{3-7b}\end{eqnarray}
For the weak coupling, it is sufficient to consider the equations
$C_{1}=0$ and $C_{2}=0$. They have the form
\begin{eqnarray}
 &&
 B_{1}(\textbf{p}_{1},\textbf{p}_{2})\frac{2m}{\hbar^{2}}[E(\textbf{p}_{1})+E(\textbf{p}_{2})+\delta E
 -E_{1}(\textbf{p}_{1}+\textbf{p}_{2})]=\nonumber \\ &=&2\left [b_{1}(\textbf{p}_{1})b_{1}(\textbf{p}_{2})\textbf{p}_{1}\textbf{p}_{2}-
 p_{1}^{2}b_{1}(\textbf{p}_{1})b_{2}(\textbf{p}_{1};\textbf{p}_{2})-
 p_{2}^{2}b_{1}(\textbf{p}_{2})b_{2}(\textbf{p}_{2};\textbf{p}_{1}) \right ]
 -\nonumber \\ &-&\frac{1}{N}\sum\limits_{\textbf{q}\neq 0}^{\textbf{q}+\textbf{p}\neq 0}
  B_{2}(\textbf{q};\textbf{p}_{1},\textbf{p}_{2})\textbf{q}(\textbf{q}+\textbf{p})-
 \frac{1}{N}\sum\limits_{\textbf{q}\neq 0}
  B_{3}(\textbf{q},-\textbf{q};\textbf{p}_{1},\textbf{p}_{2})\textbf{q}^{2},
       \label{3-8}\end{eqnarray}
\begin{eqnarray}
 && B_{2}(\textbf{q};\textbf{p}_{1},\textbf{p}_{2})\frac{2m}{\hbar^{2}}[E(\textbf{p}_{1})+E(\textbf{p}_{2})+\delta
 E-E_{1}(\textbf{q})-E_{1}(\textbf{q}+\textbf{p})]=\nonumber \\ &=&
  -b_{2}(\textbf{q};\textbf{p}_{1})b_{2}(\textbf{q}+\textbf{p}_{1};\textbf{p}_{2})(\textbf{q}+\textbf{p}_{1})^{2}
  -b_{2}(-\textbf{q}-\textbf{p};\textbf{p}_{1})b_{2}(-\textbf{q}-\textbf{p}_{2};\textbf{p}_{2})(\textbf{q}+\textbf{p}_{2})^{2}-\nonumber
  \\&-&b_{2}(\textbf{q};\textbf{p}_{2})b_{2}(\textbf{q}+\textbf{p}_{2};\textbf{p}_{1})(\textbf{q}+\textbf{p}_{2})^{2}
  -b_{2}(-\textbf{q}-\textbf{p};\textbf{p}_{2})b_{2}(-\textbf{q}-\textbf{p}_{1};\textbf{p}_{1})(\textbf{q}+\textbf{p}_{1})^{2}+\nonumber
 \\ &+&[b_{2}(\textbf{q};\textbf{p}_{1})+b_{2}(-\textbf{q}-\textbf{p}_{1};\textbf{p}_{1})]b_{1}(\textbf{p}_{2})\textbf{p}_{2}(\textbf{q}+\textbf{p}_{1})
 +\nonumber \\&+& [b_{2}(\textbf{q};\textbf{p}_{2})+b_{2}(-\textbf{q}-\textbf{p}_{2};\textbf{p}_{2})]b_{1}(\textbf{p}_{1})\textbf{p}_{1}(\textbf{q}+\textbf{p}_{2})
 \label{3-9} -\\&-&
 [b_{2}(-\textbf{q}-\textbf{p};\textbf{p}_{1})+b_{2}(\textbf{q}+\textbf{p}_{2};\textbf{p}_{1})]b_{1}(\textbf{p}_{2})\textbf{p}_{2}(\textbf{q}+\textbf{p}_{2})
 -\nonumber \\&-&
 [b_{2}(-\textbf{q}-\textbf{p};\textbf{p}_{2})+b_{2}(\textbf{q}+\textbf{p}_{1};\textbf{p}_{2})]b_{1}(\textbf{p}_{1})\textbf{p}_{1}(\textbf{q}+\textbf{p}_{1})
 -\nonumber \\ &-&2p_{1}^{2}b_{1}(\textbf{p}_{1})b_{3}(\textbf{q},\textbf{p}_{1};\textbf{p}_{2})
 -2p_{2}^{2}b_{1}(\textbf{p}_{2})b_{3}(\textbf{q},\textbf{p}_{2};\textbf{p}_{1})
 -N\delta E b_{1}(\textbf{p}_{1})b_{1}(\textbf{p}_{2})
 \frac{2m}{\hbar^{2}}(\delta_{\textbf{q},-\textbf{p}_{1}}+\delta_{\textbf{q},-\textbf{p}_{2}})\nonumber
 \\ &+&2B_{1}(\textbf{p}_{1},\textbf{p}_{2})\textbf{p}
 [\textbf{q}a_{2}(\textbf{q})-(\textbf{p}+\textbf{q})a_{2}(\textbf{p}+\textbf{q})-\textbf{p}a_{3}(\textbf{p},\textbf{q})]
 -\nonumber \\ &-&\frac{1}{N} \sum\limits_{\textbf{q}_{1}\neq 0}
 B_{3}(\textbf{q}_{1},-\textbf{q}-\textbf{q}_{1}-\textbf{p};\textbf{p}_{1},\textbf{p}_{2})\textbf{q}_{1}(\textbf{q}+\textbf{q}_{1}+\textbf{p})
 +\nonumber \\ &+&\frac{1}{N}\sum\limits_{\textbf{q}_{1}\neq 0}
 B_{3}(\textbf{q}_{1},\textbf{q}-\textbf{q}_{1};\textbf{p}_{1},\textbf{p}_{2})\textbf{q}_{1}(\textbf{q}-\textbf{q}_{1})-
 \frac{1}{N}\sum\limits_{\textbf{q}_{1}\neq 0}
 B_{4}(\textbf{q}_{1},-\textbf{q}_{1},\textbf{q};\textbf{p}_{1},\textbf{p}_{2})q^{2}_{1},
       \nonumber\end{eqnarray}
where $\textbf{p}=\textbf{p}_{1}+\textbf{p}_{2}$,
$E_{1}(\textbf{q})=\frac{\hbar^{2}q^{2}}{2m}(1-2a_{2}(\textbf{q}))$,
and $\delta_{\textbf{q},-\textbf{p}}$
is the Kronecker delta. In this case,
$B_{2}(\textbf{q};\textbf{p}_{1},\textbf{p}_{2})=B_{2}(-\textbf{q}-\textbf{p};\textbf{p}_{1},\textbf{p}_{2})$.

Let us present the functions $\psi_{\textbf{p}_{1}}$,
$\psi_{\textbf{p}_{2}},$ and
$\delta\psi_{\textbf{p}_{1}\textbf{p}_{2}}$ in (\ref{3-3}) in the
form of expansions (\ref{2-4}) and  (\ref{3-4}). Then the
``leading'' term in the expansion of
$\psi_{\textbf{p}_{1}\textbf{p}_{2}}$ is
$A\rho_{-\textbf{p}_{1}}\rho_{-\textbf{p}_{2}}$. Let us write the
functions $\psi_{\textbf{p}_{1}}$, $\psi_{\textbf{p}_{2}}$  in the
form $b_{1}(\textbf{p}_{1})\tilde{\psi}_{\textbf{p}_{1}}$,
$b_{1}(\textbf{p}_{2})\tilde{\psi}_{\textbf{p}_{2}}$. Then we
present $\psi_{\textbf{p}_{1}}\psi_{\textbf{p}_{2}}$ as a series,
where the first term is
$b_{1}(\textbf{p}_{1})b_{1}(\textbf{p}_{2})\rho_{-\textbf{p}_{1}}\rho_{-\textbf{p}_{2}}$.
The corresponding terms in the expansion of
$\delta\psi_{\textbf{p}_{1}\textbf{p}_{2}}$ (\ref{3-4}) have the
form
$\frac{B_{2}(-\textbf{p}_{1};\textbf{p}_{1},\textbf{p}_{2})+B_{2}(-\textbf{p}_{2};\textbf{p}_{1},\textbf{p}_{2})}{2N^{1/2}}
 \rho_{-\textbf{p}_{1}}\rho_{-\textbf{p}_{2}}$. Eventually, the coefficient of $\rho_{-\textbf{p}_{1}}\rho_{-\textbf{p}_{2}}$
in the expansion of the function
$\psi_{\textbf{p}_{1}\textbf{p}_{2}}$ (\ref{3-3}) is
$A=b_{1}(\textbf{p}_{1})b_{1}(\textbf{p}_{2})+\frac{B_{2}(-\textbf{p}_{1};\textbf{p}_{1},\textbf{p}_{2})+B_{2}(-\textbf{p}_{2};\textbf{p}_{1},\textbf{p}_{2})}{2N}$.
Let us represent the function $\psi_{\textbf{p}_{1}\textbf{p}_{2}}$
(\ref{3-3}) in the form
$\psi_{\textbf{p}_{1}\textbf{p}_{2}}=A\tilde{\psi}_{\textbf{p}_{1}\textbf{p}_{2}}$,
where $\tilde{\psi}_{\textbf{p}_{1}\textbf{p}_{2}}=
\frac{b_{1}(\textbf{p}_{1})b_{1}(\textbf{p}_{2})}{A}
\tilde{\psi}_{\textbf{p}_{1}}\tilde{\psi}_{\textbf{p}_{2}}+\frac{\delta\psi_{\textbf{p}_{1}\textbf{p}_{2}}}{A\sqrt{N}}$.
Since the interaction of phonons is very weak, the term
$\frac{B_{2}(-\textbf{p}_{1};\textbf{p}_{1},\textbf{p}_{2})+B_{2}(-\textbf{p}_{2};\textbf{p}_{1},\textbf{p}_{2})}{2N}$
in $A$ should be less than
$b_{1}(\textbf{p}_{1})b_{1}(\textbf{p}_{2})$ by $\sqrt{N}$ or even
$N$ times. Therefore, $
\frac{b_{1}(\textbf{p}_{1})b_{1}(\textbf{p}_{2})}{A}\approx 1$. As a
result, $\tilde{\psi}_{\textbf{p}_{1}\textbf{p}_{2}}=
\tilde{\psi}_{\textbf{p}_{1}}\tilde{\psi}_{\textbf{p}_{2}}+\frac{\delta\psi_{\textbf{p}_{1}\textbf{p}_{2}}}{A\sqrt{N}}$.
Here, $\tilde{\psi}_{\textbf{p}}$ is a one-phonon function
(\ref{2-4}) with $b_{1}=1$. In this case, $b_{j\geq 2}$ satisfy the
equations from Appendix 2, in which $b_{1}=1$. Represent the term
$\delta\psi_{\textbf{p}_{1}\textbf{p}_{2}}/A$ in the form
(\ref{3-4}). Then we consider the factor $A$ to be normalizing and
include it in $A_{\textbf{p}}$ (see (\ref{2-3})). Such
transformations lead to the necessity to set
$b_{1}(\textbf{p}_{1})=b_{1}(\textbf{p}_{2})=1$ and
$B_{2}(-\textbf{p}_{1};\textbf{p}_{1},\textbf{p}_{2})=B_{2}(-\textbf{p}_{2};\textbf{p}_{1},\textbf{p}_{2})=0$
in Eqs. (\ref{3-8}), (\ref{3-9}) and the equations of Appendix 2.

We consider the coupling to be weak: $\gamma\ll 1$, but $\gamma \gg
N^{-2}$ (the latter is necessary for the linearity of the dispersion
law at small $p$). In this case, we can seek $\delta E$ and
$\delta\psi_{\textbf{p}_{1}\textbf{p}_{2}}$ in the zero
approximation. This means \cite{yuv2,yuv1} that all sums in the
chain of equations for $B_{j}$ and $\delta E$ should be neglected.
As a result, Eq. (\ref{3-8}) takes the form
\begin{eqnarray}
 B_{1}(\textbf{p}_{1},\textbf{p}_{2})=\frac{\hbar^{2}}{m}\frac{\textbf{p}_{1}\textbf{p}_{2}-
 p_{1}^{2}b_{2}(\textbf{p}_{1};\textbf{p}_{2})-
 p_{2}^{2}b_{2}(\textbf{p}_{2};\textbf{p}_{1})}{E(\textbf{p}_{1})+E(\textbf{p}_{2})+\delta E
 -E_{1}(\textbf{p}_{1}+\textbf{p}_{2})}.
       \label{3-10}\end{eqnarray}
Let us set in (\ref{3-9}) $\textbf{q}=-\textbf{p}_{1}$. Then Eq.
(\ref{3-9}) reads
\begin{eqnarray}
 && 0= -b_{2}(-\textbf{p}_{2};\textbf{p}_{1})b_{2}(\textbf{p}_{1}-\textbf{p}_{2};\textbf{p}_{2})(\textbf{p}_{2}-\textbf{p}_{1})^{2}
  - b_{2}(-\textbf{p}_{1};\textbf{p}_{2})b_{2}(\textbf{p}_{2}-\textbf{p}_{1};\textbf{p}_{1})(\textbf{p}_{2}-\textbf{p}_{1})^{2}
  \nonumber
 \\ &-&[b_{2}(-\textbf{p}_{1};\textbf{p}_{2})+b_{2}(\textbf{p}_{1}-\textbf{p}_{2};\textbf{p}_{2})]\textbf{p}_{1}(\textbf{p}_{1}-\textbf{p}_{2})
 - [b_{2}(-\textbf{p}_{2};\textbf{p}_{1})+b_{2}(\textbf{p}_{2}-\textbf{p}_{1};\textbf{p}_{1})]\textbf{p}_{2}(\textbf{p}_{2}-\textbf{p}_{1})
  \nonumber \\ &-&2p_{1}^{2}b_{3}(-\textbf{p}_{1},\textbf{p}_{1};\textbf{p}_{2})
 -2p_{2}^{2}b_{3}(-\textbf{p}_{1},\textbf{p}_{2};\textbf{p}_{1})
 -N\delta E
 \frac{2m}{\hbar^{2}}(1+\delta_{\textbf{p}_{2},\textbf{p}_{1}})\nonumber
 \\ &+&2B_{1}(\textbf{p}_{1},\textbf{p}_{2})\textbf{p}
 [-\textbf{p}_{1}a_{2}(-\textbf{p}_{1})-\textbf{p}_{2}a_{2}(\textbf{p}_{2})-\textbf{p}a_{3}(\textbf{p},-\textbf{p}_{1})].
       \label{3-11}\end{eqnarray}
Equation (\ref{3-9}) for $\textbf{q}=-\textbf{p}_{2}$ is also
reduced to (\ref{3-11}) (to sight this, one needs to consider the
relations $a_{2}(-\textbf{p})=a_{2}(\textbf{p})$,
$a_{3}(\textbf{p},-\textbf{p}_{1})=a_{3}(\textbf{p},\textbf{p}_{1}-\textbf{p})$
and
$b_{3}(\textbf{p}_{1},\textbf{p}_{2};\textbf{p}_{3})=b_{3}(-\textbf{p}_{1}-\textbf{p}_{2}-\textbf{p}_{3},\textbf{p}_{2};\textbf{p}_{3})$).
Equations (\ref{3-10}), (\ref{3-11}) allow us to find
$B_{1}(\textbf{p}_{1},\textbf{p}_{2})$ and $\delta E$. From Eq.
(\ref{3-9}) at $\textbf{q}\neq -\textbf{p}_{1}, -\textbf{p}_{2}$ we
can determine $B_{2}(\textbf{q};\textbf{p}_{1},\textbf{p}_{2})$.

Consider the case $\textbf{p}_{2}=\textbf{p}_{1}$. According to
(\ref{2-4}), $\textbf{q}$ in $b_{2}(\textbf{q};\textbf{p})$ must be
nonzero. Therefore, if  (\ref{3-11}) includes the term
$b_{2}(0;\textbf{p})$, this term should be dropped. Then relations
(\ref{3-10}), (\ref{3-11}) yield
\begin{eqnarray}
 B_{1}(\textbf{p}_{1},\textbf{p}_{1})=\frac{p_{1}^{2}[2-
 4b_{2}(\textbf{p}_{1};\textbf{p}_{1})]}{(2m/\hbar^{2})[2E(\textbf{p}_{1})+\delta E
 -E_{1}(2\textbf{p}_{1})]},
       \label{3-12}\end{eqnarray}
\begin{eqnarray}
  B_{1}(\textbf{p}_{1},\textbf{p}_{1}) = -\frac{2p_{1}^{2}b_{3}(-\textbf{p}_{1},\textbf{p}_{1};\textbf{p}_{1})
  +(2m/\hbar^{2}) N\delta E }{4p_{1}^{2}
 [a_{2}(\textbf{p}_{1})+a_{3}(2\textbf{p}_{1},-\textbf{p}_{1})]}.
       \label{3-13}\end{eqnarray}
Equations (\ref{3-12}) and (\ref{3-13}) give a square equation for
$\delta E$ with the roots
\begin{equation}
 \delta E_{\pm} = -\tilde{E}_{+}\pm
 \sqrt{\tilde{E}_{-}^{2}-8N^{-1}[\hbar^{2}p_{1}^{2}/(2m)]^{2}[1-2b_{2}(\textbf{p}_{1};\textbf{p}_{1})][a_{2}(\textbf{p}_{1})+a_{3}(2\textbf{p}_{1},-\textbf{p}_{1})]},
        \label{3-14}\end{equation}
where
\begin{equation}
 \tilde{E}_{\pm} =
 E(p_{1})-\frac{E_{1}(2p_{1})}{2}\pm b_{3}(-\textbf{p}_{1},\textbf{p}_{1};\textbf{p}_{1})\frac{\hbar^{2}p_{1}^{2}}{2mN}.
        \label{3-15}\end{equation}
At $p_{1}\rightarrow 0$ and $\gamma \ll 1,$ the formulae in Appendix
2 yield
\begin{equation}
 a_{3}(2\textbf{p}_{1},-\textbf{p}_{1})=a_{3}(\textbf{p}_{1},\textbf{p}_{1}) \approx
 -a_{2}(\textbf{p}_{1})/4, \quad
 b_{2}(\textbf{p}_{1};\textbf{p}_{1})\approx 1/8.
        \label{3-16}\end{equation}
Using the relation
$a_{4}(-\textbf{p}_{1},\textbf{p}_{1},\textbf{p}_{1}) \approx
 7a_{2}(\textbf{p}_{1})/16$ \cite{yuv2}, we get $b_{3}(-\textbf{p}_{1},\textbf{p}_{1};\textbf{p}_{1})\approx -5/32$.
Therefore, relations (\ref{3-14}), (\ref{3-15}) are reduced to
\begin{equation}
 \delta E_{\pm} = -\tilde{E}_{+}\pm
 \sqrt{\tilde{E}_{-}^{2}-\frac{9}{2N}\left (\frac{\hbar^{2}p_{1}^{2}}{2m}\right )^{2}a_{2}(\textbf{p}_{1})},
        \label{3-17}\end{equation}
\begin{equation}
 \tilde{E}_{\pm} =
 E(p_{1})-\frac{E_{1}(2p_{1})}{2}\mp \frac{5}{32N}\frac{\hbar^{2}p_{1}^{2}}{2m},
        \label{3-18}\end{equation}
where $E(p_{1})=\hbar p_{1}v_{s}$, $a_{2}(\textbf{p}_{1})\approx
-\alpha_{\textbf{p}_{1}}/2\approx
-\frac{\sqrt{m\rho\nu(p_{1})}}{\hbar p_{1}}$, and $E_{1}(2p_{1})$,
$v_{s}$ are determined by formulae (\ref{2-8}), (\ref{2-10}). At
$N\gg 1, \gamma \lsim N^{-1}$, the corrections $\frac{9}{2N}\left
(\frac{\hbar^{2}p_{1}^{2}}{2m}\right )^{2}a_{2}(\textbf{p}_{1})$ and
$\frac{5}{32N}\frac{\hbar^{2}p_{1}^{2}}{2m}$ in (\ref{3-17}),
(\ref{3-18}) are negligible, and solutions (\ref{3-17}),
(\ref{3-18}) take the simple form
\begin{equation}
 \delta E_{+} \approx 2|\tilde{E}|,  \quad  \delta E_{-} \approx
 -\frac{9E(p_{1})}{8N}\frac{\hbar^{2}p_{1}^{2}}{2m|\tilde{E}|},
        \label{3-19}\end{equation}
\begin{equation}
 \tilde{E} \approx
 E(p_{1})-\frac{E_{B}(2p_{1})}{2}.
        \label{3-20}\end{equation}
Since $\delta E_{+}>\delta E_{-}$, namely the solution $\delta
E_{-}$ should be realized in Nature. Thus, we have found the energy
of interaction, $\delta E$, of two phonons with the same momentum
$\hbar p_{1}$ at $p_{1}\rightarrow 0$ and weak coupling
($N^{-2}\ll\gamma \ll 1$). This result is new.

At the considered parameters of the system we have $|\tilde{E}|\sim
\frac{\hbar^{2}p_{1}^{2}}{2m}$. Therefore, $\delta E_{-} \sim
-E(p_{1})/N$. In this case, relations (\ref{3-13}), (\ref{3-16})
yield $B_{1}(\textbf{p}_{1},\textbf{p}_{1})\sim
 -1$. It is natural to expect that $|B_{1}(\textbf{p}_{1},\textbf{p}_{2})|\sim
1$ also at $\textbf{p}_{2}\neq \textbf{p}_{1}$. In this case, Eq.
(\ref{3-9}) yields
$|B_{2}(\textbf{q};\textbf{p}_{1},\textbf{p}_{2})|\sim 1$. That is,
the term $\delta\psi_{\textbf{p}_{1}\textbf{p}_{2}}/\sqrt{N}$ in
formula (\ref{3-3}) is less than the main term
$\psi_{\textbf{p}_{1}}\psi_{\textbf{p}_{2}}$ by $\sim N$ times.
These estimates show that the interaction of two phonons is indeed
very weak.

Let us return to the question about the nature of a hole. In the above equations, we pass to a
1D point potential. Compare $\delta E_{-}$ with the
quantity
\begin{equation}
 \delta E_{h}=E_{h}(p=4\pi/L)-2E_{p}(p=2\pi/L)
        \label{3-21}\end{equation}
equal to the difference of the energy of a hole with the quantum numbers
$\{I_{i}\}=(-\frac{N-1}{2},-\frac{N-3}{2},\ldots,\frac{N-5}{2},1+\frac{N-3}{2},1+\frac{N-1}{2})$
 and two energies of a free ``particle''  (phonon) with the quantum
numbers
$\{I_{i}\}=(-\frac{N-1}{2},-\frac{N-3}{2},\ldots,\frac{N-5}{2},\frac{N-3}{2},1+\frac{N-1}{2})$.
The quantities $p=4\pi/L$ and $p=2\pi/L$ in (\ref{3-21}) are
momenta. The values of $E_{h}(p=4\pi/L)$ and $E_{p}(p=2\pi/L)$ can
be found numerically from the Yang--Yang's  equations (\ref{2}) and
formulae (\ref{2-14}), (\ref{2-15}). The value of $\delta E_{-}$
follows from Eqs. (\ref{3-17}) and (\ref{3-18}), where we set
$\nu(p)=2c$, $\hbar=2m=1$, $c/\rho=\gamma$, and $ p_{1}=2\pi/L$.

\begin{figure}[ht]
\centerline{\includegraphics[width=85mm]{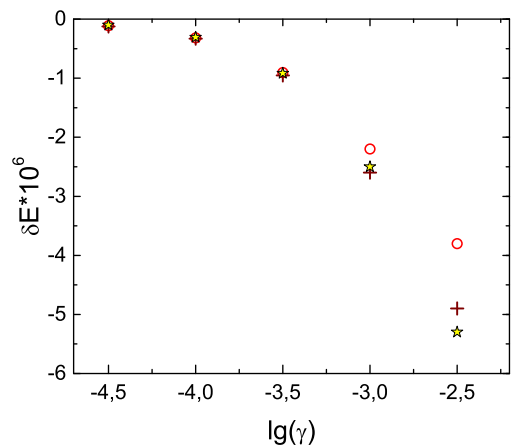} } \caption{
[Color online] Functions $\delta E_{-}(\gamma)$ (\ref{3-17}),
(\ref{3-18})  (circles),  $\delta E_{-}(\gamma)$ (\ref{3-19}),
(\ref{3-20}) (crosses), and  $\delta E_{h}(\gamma)$ (\ref{3-21})
(stars); $\rho=1$, $N=1000$. All values of $\delta E$ are multiplied
by $10^{6}$.
 \label{fig2}}
\end{figure}

It is seen from Fig. 2 that the energy of interaction of two phonons
($\delta E_{-}$) is close to $\delta E_{h}$, if $N^{-2}\ll\gamma
\lsim N^{-1}$. The very small value of $\delta E_{h}$ is an
indicator of the nature of a hole. The closeness of the values of
$\delta E_{-}$ and $\delta E_{h}$ proves that the hole
$\{I_{i}\}=(-\frac{N-1}{2},-\frac{N-3}{2},\ldots,\frac{N-5}{2},1+\frac{N-3}{2},1+\frac{N-1}{2})$
coincides with two interacting phonons, each characterized by the
collection
$\{I_{i}\}=(-\frac{N-1}{2},-\frac{N-3}{2},\ldots,\frac{N-5}{2},\frac{N-3}{2},1+\frac{N-1}{2})$.

In the region $N^{-1}\ll\gamma \ll 1$ the quantities $\delta E_{-}$
and $\delta E_{h}$ are considerably different, since we found a
solution for $\delta E_{-}$ only in  zero approximation. The error
of the numerical calculation of $\delta E_{h}$ should also be
significant in this case.

We note that, to obtain namely a two-phonon solution, it is
necessary firstly to set the orders of the quantities $B_{j}$ and
$\delta E$. Otherwise, we can arrive at another solution, since the
function $\psi_{\textbf{p}_{1}\textbf{p}_{2}}$ (\ref{3-3}),
(\ref{3-4}) can describe \textit{any} excited state with the total
momentum $\hbar(\textbf{p}_{1}+\textbf{p}_{2})$ (see Appendix 1). We
took the two-phonon nature of the state into account with the help
of the condition
$|[B_{2}(-\textbf{p}_{1};\textbf{p}_{1},\textbf{p}_{2})+B_{2}(-\textbf{p}_{2};\textbf{p}_{1},\textbf{p}_{2})]/(2N)|\ll
|b_{1}(\textbf{p}_{1})b_{1}(\textbf{p}_{2})|$.

The above two-phonon solution should be contained in Eqs.
(\ref{9-1})--(\ref{9-6}) of Appendix 2, since any (not only
one-phonon) excited state of the system with the total momentum
$\textbf{p}$ is described by the function $\psi_{\textbf{p}}\Psi_0$
(\ref{2-3}), (\ref{2-4}) (see Sect. 2). In other words, the
Vakarchuk--Yukhnovskii's equations (see Appendix 2) contain
solutions for all excited state of a Bose gas.

We have shown that, at $j = 1, 2$ the hole with the momentum $j 2\pi
/L$ is $j $ interacting phonons $(0,\ldots,0,1)$. Let, for some $j =
J > 2,$ this assertion be wrong. Since the solution of the
Lieb--Liniger equations is unique \cite{takahashi1999}, the system
of point bosons would not contain the state with $J$ interacting
phonons $(0,\ldots,0,1)$. This is strange from the physical point of
view and would lead to the violation of the continuous transition
from solutions for nonpoint bosons to solutions for point ones.
However, such transition should exist \cite{seiringer2008}. Hence,
for all $j = 1, \ldots, N,$ the hole with the momentum $j 2\pi /L$
is $j $ interacting phonons $(0,\ldots,0,1)$.

\section{Additional arguments.}
Consider a 1D Bose gas with point interaction. Let us find the limit
$c\rightarrow 0$ for the Lieb--Liniger solutions
\cite{gaudinm,ll1963}
\begin{equation}
 \psi_{\{k \}}(x_{1},\ldots,x_{N})=const \cdot\sum\limits_{P}a(P)e^{i\sum\limits_{l=1}^{N} k_{P_{l}}x_{l}},
      \label{4-1} \end{equation}
\begin{equation}
 a(P)=\prod\limits_{j<l}\left (1+\frac{ic}{k_{P_{j}}-k_{P_{l}}} \right
 ).
      \label{4-2} \end{equation}
For the state $\{n_{i}\}=(0,\ldots,0,1)$,  at $c\rightarrow 0$ we
get $\{k_{i}\}=(0,\ldots,0,2\pi/L)$. Relations (\ref{4-1}) and
(\ref{4-2}) yield $a(P)=1$ and
\begin{equation}
 \psi_{\{k \}}\equiv \psi_{1}=c_{1}\rho_{-k_{N}},
      \label{4-3} \end{equation}
where $k_{N}=2\pi/L$. For the state $\{n_{i}\}=(0,\ldots,0,1,1),$ we
get $\{k_{i}\}\approx (0,\ldots,0,2\pi/L,2\pi/L)$. Then relations
(\ref{4-1}), (\ref{4-2}) yield $a(P)=1$ and
\begin{equation}
 \psi_{\{k \}}\equiv \psi_{11}=c_{11}\left
 (\rho_{-k_{N}}\rho_{-k_{N}}-\frac{\rho_{-2k_{N}}}{\sqrt{N}}\right ).
      \label{4-4} \end{equation}
Here, while calculating $a(P),$ we take into account that
$(k_{N}-k_{N-1})|_{c\rightarrow 0}\sim c^{1/2}$. Functions
(\ref{4-3}) and (\ref{4-4}) coincide with the wave functions of a
system of free bosons, in which one or two (respectively) atoms have
the momentum $2\pi/L$. The normalizing coefficients are
$c_{1}=L^{-N/2}$, $c_{11}=\sqrt{\frac{N}{N-1}}c_{1}$ \cite{yuv1}.
Since $|\rho_{-k_{N}}|\sim 1$ for the overwhelming majority of
configurations $(x_{1},\ldots,x_{N})$, the comparison of $\psi_{11}$
(\ref{4-4}) and $\psi_{1}$ (\ref{4-3}) shows that in the limit
$c\rightarrow 0$ \textit{the hole $(0,\ldots,0,1,1)$ is two
interacting particles $(0,\ldots,0,1)$}, which agrees with the
result of the previous section. At $c=0$ we have, of course, free
atoms instead of quasiparticles.

The one-phonon and two-phonon solutions (\ref{2-3}) and (\ref{3-3})
pass at $c= 0$ to solutions (\ref{4-3}) and (\ref{4-4}). To
demonstrate this with the formulae in Sections 2 and 3, we take the
relations $a_{j}=0$, $b_{j\geq 2}=0$, $B_{j\geq 2}=0$, and $\delta
E=0$ into account. Relation (\ref{3-12}) yields
$B_{1}(p_{1},p_{1})=-1$. Thus, Eqs. (\ref{2-3}), (\ref{2-4}), and
(\ref{3-3}) describe free bosons at the zero interaction and phonons
at a nonzero one (if the interaction is switched-on, the functions
$\psi_{\textbf{p}_{1}}$, $\psi_{\textbf{p}_{1}\textbf{p}_{2}}$ vary
negligibly, but the dispersion law $E(p)\sim p^{2}$ transits into
$E(p)\approx v_{s}p$ due to a change of $\Psi_{0}$).

It is clear that any Lieb--Liniger solution (\ref{4-1}) can be
presented in the form (\ref{2-3}), (\ref{2-4}). It would be of
interest to get solutions (\ref{2-3}), (\ref{2-4}), and (\ref{3-3})
directly from (\ref{4-1}) at $c\neq 0$. This is a task for the
future.

Both in the Gaudin's numbering and in the collective variables
method, each excited state of a 1D system is described by the
collection of quantum numbers $\{n_{i}\}\equiv
(n_{1},n_{2},\ldots,n_{N})$ corresponding to the collection of
quasiparticles with the momenta $p_{1},\ldots,p_{N}$, where
$p_{j}=2\pi n_{j}/L$. That is, there is one-to-one correspondence
between solutions in the collective variables  method at $\nu(p)=
2c$ and solutions in the Lieb--Liniger approach. In this case, the
uniqueness of a solution for each collection $\{n_{i}\}$ was proved
only for the Lieb--Liniger approach \cite{takahashi1999}.

The calculation of the statistical sum of a 1D system of point
bosons at $N=\infty$ gives \cite{mt2015}
\begin{equation}
F|_{T\rightarrow 0}= E_{0} +k_{B}T\sum\limits_{l=\pm 1, \pm 2,
\ldots}\ln{\left (1-e^{-\frac{E_{p}(p_{l})}{k_{B}T}}\right )},
     \label{4-5} \end{equation}
where $E_{p}(p_{l})$ is the dispersion law of particles.  The
calculation \cite{mt2015} involves \textit{all} states of the system
(including the ground state, particles, and holes). Formula
(\ref{4-5}) is exact at $N=\infty$ and $T \rightarrow 0$. Equation
(\ref{4-5}) is the known formula for the free energy of an ensemble
of noninteracting Bose quasiparticles. The verification
\cite{mtjltp2017} indicates that formula (\ref{4-5})  and the
Yang--Yang approach \cite{yangs1969} lead to identical thermodynamic
solutions $F, S$. If we consider formally the state
$\{n_{i}\}=(0,\ldots,0,1)$ as a hole, then any excited state
$(n_{1},\ldots,n_{N})$ can be approximately considered as a
collection of noninteracting holes. This leads again to formula
(\ref{4-5}) with the replacement of $E_{p}(p)$ by the dispersion law
of holes $E_{h}(p)$.  The analysis of the present work shows that
such dualism of holes and particles is, apparently, physical at a
strong coupling, but is illusory at the weak coupling.

The analysis of Sections $1$--$4$  shows  that, at the weak
coupling, the hole is merely  a collection of identical interacting
phonons with the momentum $\pm 2\pi /L$. This corresponds to the
Gaudin's numbering (see Eq. (\ref{1})). Therefore, the introduction
of quasiparticles with the help of Gaudin's numbering
\cite{mtjltp2017,mt2015} is more physical, at least at the weak
coupling. In this case, the curve of holes $E_{h}(p)$ describes the
excited states with minimum energy for given $p$. The Yang--Yang's
numbering (see Eq. (\ref{2})) is also useful: using it, it is easy
to find the energy of quasiparticles at a strong coupling.

We recall also the arguments by Feynman
\cite{fey1954,fc1956,fey1972}. According to them, only the single
dispersion law, corresponding to phonons, should be in the region of
small $E, p$. Such conclusion is in agreement with our analysis.

\section{Regime of infinitely strong repulsion and experiments}
Above, we studied the regime of the weak coupling. We now consider
the Tonks--Girardeau gas: $\gamma =+\infty$ (see reviews
\cite{bloch2008,cazalilla2011} and  experimental works
\cite{paredes2004,nagerl2015,weiss2020}). This regime is the most
unusual. In this case, the point bosons are  impenetrable.
Therefore, two bosons cannot stay at a single point, which is
similar to fermions. As a result, the system of bosons acquires some
fermionic properties \cite{lieb1963,girardeau}. In particular, the
energy levels coincide with those of a system of free fermions. It
is interesting, since any perturbation of a system of interacting
bosons is a collection of oscillatory modes. Hence, for $\gamma =
+\infty$, the oscillatory modes reproduce exactly the energy levels
of the excited system of free fermions.

For a 1D system of impenetrable bosons, Girardeau obtained the
dispersion law \cite{girardeau}
\begin{equation}
  E(p)=\frac{\hbar^{2}p^{2}}{2m}+\frac{\hbar^{2}|p|\pi \rho}{m}\frac{N-1}{N}
  \label{st-1}     \end{equation}
that is a limiting case of the Lieb's dispersion law of
``particles'' \cite{lieb1963}. The wave function (WF) of any state
$(n_{1},\ldots,n_{N})$ can be represented as
 \begin{equation}
    \Psi_{p}(x_{1},\ldots ,x_{N}) =
  \psi_{p}\Psi_{0}.
  \label{st-3}     \end{equation}
If $n_{j}\geq 0$ for any $j$, then $\psi_{p}$  is the Schur function
$S_{\lambda}$ \cite{macdonald}:
 \begin{equation}
   S_{\lambda}(z_{1},\ldots ,z_{N}) = \det{(h_{\lambda_{i}-i+j})}_{1\leq i,j \leq l_{1}},
   \quad l_{1}\geq \lambda_{1}^{\prime},
    \label{m-1}     \end{equation}
 \begin{equation}
   S_{\lambda}(z_{1},\ldots ,z_{N}) = \det{(e_{\lambda^{\prime}_{i}-i+j})}_{1\leq i,j \leq l_{2}},
   \quad l_{2}\geq \lambda_{1}.
    \label{m-2}     \end{equation}
Formulae (\ref{m-1}) and (\ref{m-2}) are equivalent. Here,
$z_{q}=e^{ik_{1}x_{q}}$, $ k_{1}=2\pi/L$, $h_{j}$ and $e_{j}$ are
the $j$th complete symmetric function and the $j$th elementary
symmetric function, respectively \cite{macdonald}:
\begin{equation}
h_{j}(z_{1},\ldots ,z_{N})=\sum\limits_{s_{1}\leq
s_{2}\leq\ldots\leq s_{j}}z_{s_{1}}\cdots z_{s_{j}},
  \label{m-3}  \end{equation}
 \begin{equation}
 e_{j}(z_{1},\ldots ,z_{N})=\sum\limits_{s_{1}< s_{2}<\ldots < s_{j}}z_{s_{1}}\cdots z_{s_{j}},
  \label{m-4}     \end{equation}
where $j\geq 0$, $e_{0}=h_{0}=1$, $h_{1}=e_{1}$ (in all similar sums
in this section, we sum over $s_{1},\ldots,s_{j}=1,\ldots,N$). The
partitions $\lambda=(\lambda_{1},\lambda_{2},\ldots)$ and
$\lambda^{\prime}=(\lambda^{\prime}_{1},\lambda^{\prime}_{2},\ldots)$
are defined in \cite{macdonald}. In particular, for the particle
$(0,\ldots,0,j)$ we have $\lambda_{1}=j, \lambda_{i\geq 2}=0$,
$\lambda^{\prime}_{1}=1,\ldots,\lambda^{\prime}_{j}=1,
\lambda^{\prime}_{i>j}=0$. In this case, Eq. (\ref{m-1}) leads to
the Girardeau's solution \cite{girardeau}
 \begin{equation}
   \psi_{p}=S_{\lambda} = h_{j}(z_{1},\ldots ,z_{N}).
       \label{m-5}     \end{equation}
For the hole $(0,\ldots,0,1,\ldots,1)$, where $1$ is repeated  $j$
times, we have $\lambda_{1}=1,\ldots,\lambda_{j}=1,
\lambda_{i>j}=0$, $\lambda^{\prime}_{1}=j, \lambda^{\prime}_{i\geq
2}=0$. Eq. (\ref{m-2}) gives  the solution
 \begin{equation}
   \psi_{p}=S_{\lambda} = e_{j}(z_{1},\ldots ,z_{N}).
       \label{m-6}     \end{equation}
This formula can be easily verified for $j=N$:
 \begin{equation}
 \psi_{p}=\sum\limits_{s_{1}< s_{2}<\ldots < s_{N}}z_{s_{1}}\cdots z_{s_{N}}=z_{1}z_{2}\cdots z_{N}=\exp{[ik_{1}(x_{1}+\ldots +x_{N})]}.
  \label{st-8}     \end{equation}
The same solution follows from the Lieb--Liniger formulae for any
$\gamma \geq 0$. This solution is also true for a nonpoint
interatomic potential of the general form. Solution (\ref{st-8})
describes the translational motion of a system with the velocity
$v=\hbar 2\pi/(mL)$. Interestingly, the solution for $N$ interacting
phonon-holes $(0,\ldots,0,1)$ coincides with (\ref{st-8}).

We will understand the properties of a system better, if we will
determine the structure of WFs for the lowest states.  The WF of the
particle $(0,\ldots,0,1)$ is \cite{girardeau}
 \begin{equation}
    \Psi_{p}(x_1,\ldots ,x_N) =
  \sqrt{N}\rho_{-k_{1}}\Psi_{0}.
  \label{st-2}     \end{equation}
The wave function (\ref{st-2}) corresponds to WF (\ref{2-3}),
(\ref{2-4}) of a phonon with $b_{j\geq 2}=0$. It was mentioned in
Introduction that the state $(0,\ldots,0,1)$ can be formally
considered as a particle and as a hole. The analysis in Sect. 2
indicates that at $\gamma \ll 1$ the state $(0,\ldots,0,1)$ is a
phonon. However, at $\gamma = \infty$ the energy levels of the
system coincide with those of a system of free fermions. Therefore,
the analogy with a hole becomes also physical. That is, at $\gamma =
\infty$  the state $(0,\ldots,0,1)$ can be considered as a
phonon-hole.

For the particle $(0,\ldots,0,j)$ with $j=2$  and $j=3$ we get,
respectively,
 \begin{equation}
 \psi_{p}=\sum\limits_{s_{1}\leq s_{2}}z_{s_{1}}z_{s_{2}}=\frac{N}{2!}\left (\frac{\rho_{-2k_{1}}}{\sqrt{N}}+ \rho^{2}_{-k_{1}} \right
 ),
  \label{st-5}     \end{equation}
 \begin{equation}
 \psi_{p}=\sum\limits_{s_{1}\leq s_{2}\leq s_{3}}z_{s_{1}}z_{s_{2}}z_{s_{3}}=\frac{N^{3/2}}{3!}\left (\frac{2\rho_{-3k_{1}}}{N}+
 \frac{3}{\sqrt{N}}\rho_{-k_{1}}\rho_{-2k_{1}}+ \rho^{3}_{-k_{1}} \right
 ).
  \label{st-6}     \end{equation}

For the holes with the momenta $p=4\pi/L$ and $p=6\pi/L,$ we obtain,
respectively,
 \begin{equation}
 \psi_{p}=\sum\limits_{s_{1}< s_{2}}z_{s_{1}}z_{s_{2}}=\frac{N}{2!}\left (-\frac{\rho_{-2k_{1}}}{\sqrt{N}}+ \rho^{2}_{-k_{1}} \right
 ),
  \label{st-10}     \end{equation}
   \begin{equation}
 \psi_{p}=\sum\limits_{s_{1}< s_{2}< s_{3}}z_{s_{1}}z_{s_{2}}z_{s_{3}}=\frac{N^{3/2}}{3!}\left (\frac{2\rho_{-3k_{1}}}{N}-
 \frac{3}{\sqrt{N}}\rho_{-k_{1}}\rho_{-2k_{1}}+ \rho^{3}_{-k_{1}} \right
 ).
  \label{st-11}     \end{equation}

We did not verify the normalization for formulae
(\ref{m-1})--(\ref{st-11}). If $N \gg 1,$ we have $|\rho_{-k}|\sim 1
$ for the vast majority of configurations $(x_{1},\ldots,x_{N})$.
Therefore, at $N \gg 1$ the last term dominates in solutions
(\ref{st-5}), (\ref{st-6}), (\ref{st-10}), and (\ref{st-11}), and
the remaining terms are small corrections. We can conclude that the
states $(0,\ldots,0,2)$ and $(0,\ldots,0,1,1)$ [formulae
(\ref{st-5}), (\ref{st-10})] correspond to two interacting
quasiparticles $(0,\ldots,0,1)$ (\ref{st-2}). In Sect. 3 we
considered the two-phonon state with similar structure. The states
$(0,\ldots,0,3)$ and $(0,\ldots,0,1,1,1)$ [formulae (\ref{st-6}),
(\ref{st-11})] correspond to three interacting quasiparticles
$(0,\ldots,0,1)$.

To pass from the solutions $\psi_{p}$  (\ref{m-5}), (\ref{m-6}) with
the momentum $p>0$ to solutions with $p<0$, it is sufficient to
change $k_{i}\rightarrow -k_{i}$ for all $i$ in the Lieb--Liniger
equations (\ref{1}) written for a state with $p<0$. Therefore, the
solution $\psi_{p}$ for a hole $(-1,\ldots,-1,0,\ldots,0)$ and for a
particle $(-j,0,\ldots,0,0)$ can be found, by replacing
$z_{s_{i}}\rightarrow 1/z_{s_{i}}$ for all $i=1,\ldots,j$ in Eqs.
(\ref{m-3})--(\ref{st-11}). We can also use the relation
 \begin{equation}
 \psi_{-|p|}\psi_{p_{max}}=\psi_{p_{max}-|p|} \quad \quad
 (p_{max}=N2\pi/L=2\pi \rho)
  \label{st-9}     \end{equation}
which follows from the representation of WF in the form of a
determinant \cite{girardeau}; $\psi_{p_{max}}$ is given by
(\ref{st-8}). From whence, we get that the states $(-2,0,\ldots,0)$
and $(-1,-1,0,\ldots,0)$ correspond to two interacting
quasiparticles $(-1,0,\ldots,0)$, and the states $(-3,0,\ldots,0)$,
$(-1,-1,-1,0,\ldots,0)$ correspond to three interacting
quasiparticles $(-1,0,\ldots,0)$. It is natural to expect that the
particles and holes with higher momenta $|p|$ can also be considered
as a collection of interacting quasiparticles $(0,\ldots,0,1)$ or
$(-1,0,\ldots,0)$.

Such properties of particles are surprising, because,  at the weak
coupling, the particles with small $|p|$ correspond to phonons and
are indivisible structures. However, at $\gamma=\infty$ all
particles turn out to be composite structures, except for phonons
$(0,\ldots,0,1)$ and $(-1,0,\ldots,0)$. This means that each state
$(n_{1},\ldots,n_{N})$ can be, apparently, considered as a
collection of interacting phonon-holes $(0,0,\ldots,0,1)$ and (or)
$(-1,0,\ldots,0,0)$. In this case, \textit{in the Tonks--Girardeau
gas there are only two primary indivisible excitations:  the
phonon-holes $(0,0,\ldots,0,1)$ and $(-1,0,\ldots,0,0)$}. Such
system is not characterized by any dispersion law. Therefore, the
function $S(k=const,\omega)$  should not have a sharp peak. This
agrees with the theory \cite{cherny2006} according to which
$S(k=const,\omega)=const$ for $\gamma=\infty$. The experiment
\cite{nagerl2015} and the theory \cite{caux2014} testify to a
widening of the peak of the function $S(k=const,\omega),$ as
$\gamma$ increases.

In this case, at $\gamma = \infty, T\rightarrow 0$  particles and
holes interact weakly between themselves  and, therefore, are
``good'' quasiparticles (according to Landau's arguments
\cite{landau1941}). Using Eqs. (\ref{2}) with $c\rightarrow \infty$
and Eqs. (\ref{2-14}), (\ref{2-15}), we find that the energy of
interaction of two particles with the momentum $p=j2\pi/L> 0$ is
equal to $\frac{-2E_{p}}{N+j-1}$ (where $E_{p}$ is the energy of one
particle with the momentum $j2\pi/L$), and the energy of interaction
of two holes with the same momentum is $\frac{2E_{h}}{N-j+1}$ (where
$E_{h}$ is the energy of a hole with $p=j2\pi/L$).

Interestingly, the thermodynamic velocity of sound coincides with
the microscopic one found from Eq. (\ref{st-1}):
$v_{s}^{therm}=v_{s}^{mic}$ \cite{girardeau}. Therefore, Girardeau
concluded that the low-lying excitations with energy (\ref{st-1})
correspond to phonons \cite{girardeau}. However, the equality
$v_{s}^{therm}=v_{s}^{mic}$ holds also for holes (because
$E_{h}(p)=-\frac{\hbar^{2}p^{2}}{2m}+\frac{\hbar^{2}|p|\pi
\rho}{m}\frac{N+1}{N}$ \cite{lieb1963}). The curves for holes and
particles coincide at the points $p=2\pi/L$ and $p=-2\pi/L$
corresponding to the states $(0,0,\ldots,0,1)$ and
$(-1,0,\ldots,0,0)$. Therefore, the equality
$v_{s}^{therm}=v_{s}^{mic}$ indicates, apparently, only that the
excitations $(-1,0,\ldots,0,0)$ and $(0,0,\ldots,0,1)$ are phonons.
The remaining excitations may not be single phonons.

Since the levels of a system of impenetrable bosons coincide with
those of free fermions, the creation of a hole (particle) is
equivalent to a change in the momentum of one atom by the value of
the momentum of a hole (particle).  In view of this, the scattering
of an external atom on the system can be considered as the
scattering on a single atom (but not as the creation of a set of
phonon-holes $(0,\ldots,0,1)$). It is an individual process. The
probability of such process is greater than that of multiple
processes. The calculations \cite{caux2014,caux2007} show that at
$\gamma \gg 1$ the peak of the function $S(k=const,\omega)$ should
be located between the dispersion curves of particles and holes.
This result agrees with the experiment \cite{nagerl2015}.

At the weak coupling, the phonons are described by WF (\ref{2-3}),
(\ref{2-4}) corresponding to a structureless object, and a hole is a
collection of phonons $(0,\ldots,0,1)$. Therefore, the creation of a
hole cannon be considered as a change in the momentum of a single
atom, but it should be considered as a multiple process. The
probability of such processes is very small. Due to this, the peak
of $S(k,\omega)$ should be close to the dispersion curve of
particles, which is in agreement with the theory
\cite{caux2014,caux2007,glazman2008,golovach2009} and experiment
\cite{nagerl2015}.

We note that if the system contains two independent types of
excitations, then the atom flying through the system can
independently create excitations of both types. As a result, the
function $S(k=const,\omega)$ should be characterized by two peaks.
However, only one peak was observed in the experiment
\cite{nagerl2015} (even at high $p$, for which $E_{h}(p)$ and
$E_{p}(p)$ differ significantly from each other). This means that
the system contains the elementary excitations of only one type.

In the experiment \cite{inguscio2015}, the profile of
$S(k=const,\omega)$ was measured for $N\simeq 30$ atoms in  a $1D$
trap with $\gamma\approx 1$. The experiment agrees better with the
theory based on the Bethe ansatz, than with the Bogolyubov theory
\cite{bog1947,bz1955}. This is not surprising, because the
Bogolyubov formulae work at $\gamma= 1$ \cite{lieb1963}, only if $N$
is large: $N\gsim 100$ and $N\gsim 1000$ for periodic and zero
boundary conditions, respectively \cite{mt2015,mtspectr}. Moreover,
the Bogolyubov approach does not account for the interaction of
quasiparticles. However, this interaction is important for the
many-quasiparticle states, which make a significant contribution to
$S(k,\omega)$. In this case, the Bethe equations work at small $N$
and involve the interaction of quasiparticles. Nevertheless, the
approaches by Lieb and Bogolyubov are equivalent at the weak
coupling and $N\gg 1$, as shown in the present work.

Note also that the atoms were modeled in the experiments
\cite{nagerl2015,inguscio2015} as point ones, though the real atoms
have nonzero radii of interaction. The account for a finite size of
atoms can explain qualitatively, for small $k$ and $\gamma \lsim 1$,
the experimental shift of a peak of $S(k,\omega),$ as $\gamma$
increases \cite{nagerl2015}. It can be seen qualitatively, by using
the Bogolyubov formula $E_{B}(p)=\sqrt{\left
(\frac{\hbar^{2}p^{2}}{2m}\right
)^{2}+2\rho\nu(p)\frac{\hbar^{2}p^{2}}{2m}}$ for the point and
nonpoint atoms. The consideration of nonpoint atoms will lead also
to a shift of the Bogolyubov peak of $S(k=const,\omega)$ to lower
$\omega$ for the experiment \cite{inguscio2015}, which can improve
the agreement with the experimental peak.

On the whole, the properties of the Tonks--Girardeau gas are strange
and not visual.

\section{Elementary excitations}
The known  physical laws are, in fact, the simplest ways to describe
the complex connections in Nature. Therefore, it is natural to
introduce elementary excitations \cite{lieb1963} so that the
description of the properties of systems in their language be the
simplest. This implies that every state of the  system must uniquely
correspond to some collection of elementary excitations (and it
should match the structure of wave functions). Moreover,  the
interaction between excitations should become weak at $T\rightarrow
0$ \cite{landau1941}.

Any state of a system of point bosons corresponds uniquely to a
collection of quasiparticles in two classifications: if each
quasiparticle is considered as a collection of ``particles'' and if
each quasiparticle is represented as a collection of ``holes''. If
the particles and holes are introduced jointly, then the ambiguity
arises for the majority of states. For example, the state
$(n_{1},\ldots,n_{N})=(0,\ldots,0,2,2)$ can be considered as two
interacting phonons $(0,\ldots,0,0,2)$ or as two interacting holes
$(0,\ldots,0,1,1)$.

1) \textbf{Weak coupling.} Consider the state $(0,\ldots,0,2)$.
Since the states like $(0,\ldots,0,j)$ correspond to the Bogolyubov
dispersion law, we identify them with phonons. In the ``only
particles'' language, the state $(0,\ldots,0,2)$ is a particle. In
the ``only holes'' language, the state $(0,\ldots,0,1)$  should be
effectively considered as a hole, and the state $(0,\ldots,0,2)$
should be considered as two interacting holes $(0,\ldots,0,1)$.
However, WF of the state $(0,\ldots,0,2)$ is known. It is function
(\ref{2-3}), (\ref{2-4}) with $\textbf{p}=\textbf{i}_{x}4\pi/L$. If
the state $(0,\ldots,0,2)$ would be two holes $(0,\ldots,0,1)$, then
the solution for $(0,\ldots,0,2)$ would coincide with one of the
two-phonon solutions found in Sect. 3. But this is not the case: the
state $(0,\ldots,0,2)$ has a different energy, and WF (\ref{2-4})
with $\textbf{p}=\textbf{i}_{x}4\pi/L$ corresponds to one phonon and
is different by structure from the two-phonon WF
$\psi_{\textbf{p}_{1}\textbf{p}_{1}}=
\psi_{\textbf{p}_{1}}\psi_{\textbf{p}_{1}}+\frac{\delta\psi_{\textbf{p}_{1}\textbf{p}_{1}}}{\sqrt{N}}$
(\ref{3-3}) with $\textbf{p}_{1}=\textbf{i}_{x}2\pi/L$.

This shows that, at the weak coupling, the particles
$(0,\ldots,0,j)$ are indivisible elementary excitations. In this
case, the language of holes can be used only formally, for example,
in the calculation of a partition function (see Sect. 4) or the
dynamical structural factor (DSF) $S(k,\omega)$.

It is worth to note that, in the hole approach, the statistics of
holes turns out to be contradictory. The partition function leads to
formula (\ref{4-5}) with the replacement $E_{p}(p)\rightarrow
E_{h}(p)$. Such formula corresponds to the Bose statistics. But WF
has no corresponding symmetry. For example, we saw that WF of the
state $(0,\ldots,0,2)$ cannot be presented as WF of two holes
$(0,\ldots,0,1)$. Therefore, if the state $(0,\ldots,0,2)$ is
formally considered as two holes $(0,\ldots,0,1)$, then these two
holes cannot be permuted. Such discrepancy between the thermodynamic
formulae and the symmetry of WFs testifies that the partition into
quasiparticles is not consistent with the structure of WFs and,
therefore, is not quite physical. In the particle-based approach, no
such problem arises.

If we consider the particles and holes jointly, then, for each state
$(n_{1},\ldots,n_{N}),$ we should indicate the rule, according to
which this state is separated into holes and particles. In the phase
space of numbers $(n_{1},\ldots,n_{N}),$ such rule sets the boundary
between holes and particles. If the rule is formulated, then we can
calculate the partition function and DSF and can indicate which
contribution is given by holes or particles. If such rule is not
set, then the contributions of  holes and particles are unknown. In
this case, holes and particles are not defined, and nothing can be
said about their statistics. At the joint consideration of holes and
particles the Bose statistics for quasiparticles should be violated
(e.g., if we consider the state $(0,\ldots,0,1,1)$ as a hole, then
the state with two phonons $(0,\ldots,0,1)$ is lost, which violates
the properties of Bose quasiparticles). In this case, the statistics
of quasiparticles can acquire the fermionic character. That is,
establishing the boundary between holes and particles in the space
of numbers $(n_{1},\ldots,n_{N})$ is a significant point. However,
we did not see the articles, where such boundary is introduced at
the joint consideration of holes and particles.

Some of the above mentioned properties and difficulties have already
been discussed in work \cite{lieb1963}. Taking into account the
fermionic properties of a Bose system at $\gamma \gg 1$, Lieb
conserved the symmetry between particles and holes at any $\gamma $
and considered them jointly  \cite{lieb1963}. But the above analysis
shows that, at the weak coupling, this symmetry is broken to the
favor of particles.

2) \textbf{Strong coupling:} $\gamma = +\infty$. By the analysis in
the previous section, at $\gamma = +\infty$  each state can be,
apparently, considered as a collection of interacting elementary
excitations $(0,\ldots,0,1)$ and (or) $(-1,0,\ldots,0)$. In this
case, it is worth talking about the bosonic or fermionic properties
of these two indivisible excitations only. These are bosons. It
follows from the fact that one state of the system can contain
several such excitations, as is seen from solutions (\ref{st-5}),
(\ref{st-6}), (\ref{st-10}), (\ref{st-11}). Moreover, the state
$(-1,0,\ldots,0,1)$ corresponds to
\begin{equation}
 \psi_{p}=\frac{S_{\lambda}}{e_{N}}=\frac{e_{N-1}e_{1}-e_{N}e_{0}}{e_{N}}=\sum\limits_{i,j=1}^{N}\frac{z_{i}}{z_{j}}-1=N\left (
\rho_{-k_{1}}\rho_{k_{1}}- \frac{1}{N}\right ).
  \label{st-20}     \end{equation}
(To obtain this formula, one needs to extract the multiplier $z_{j}$
from $j$th row of the Slater determinant \cite{girardeau} and then
to use Eq. (\ref{m-2}) with $l_{2}=\lambda_{1}=2$,
$\lambda^{\prime}_{1}=N-1$, $\lambda^{\prime}_{2}=1$). Function
(\ref{st-20}) describes two interacting quasiparticles:
$(-1,0,\ldots,0,0)$ and $(0,0,\ldots,0,1)$.  Formula (\ref{st-20})
shows that $\psi_{p}$ is invariable under the permutation of the
quasiparticles $(-1,0,\ldots,0,0)$ and $(0,0,\ldots,0,1)$, which
corresponds to bosons.

At $\gamma = +\infty$ there is a symmetry between particles and
holes. We can describe the system in the language of particles or in
the language of holes.

\section{A hole and a soliton.}
The Lieb's hole is a stationary solution of the $N$-body
Schr\"{o}dinger equation for a cyclic system:
$\tilde{\Psi}(x_{1},\ldots,x_{N},t)=e^{-iE_{h}(p)t/\hbar}\Psi(x_{1},\ldots,x_{N})$.
This solution is characterized by a constant  density:
$\rho(x,t)=const$ \cite{sato2012}. However, the quasiclassical dark
soliton, as a solution of the 1D Gross--Pitaevskii equation, is a
solitary \textit{running density wave} of the form
$\Psi(x,t)=\Psi(x-vt)$, $\rho(x,t)=\rho(x-vt)$
\cite{tsuzuki1971,ishikawa1980}. In this case, the wave package of
one-hole states shows the properties of an immovable soliton
\cite{sato2012,sato2016,brand2018} (though the density profile
$\rho(x,t)$ of such package spreads, as $t$ increases, in contrast
to a quasiclassical soliton \cite{tsuzuki1971,ishikawa1980}).
Moreover, the conditional probability density $\rho_{N}(x)$ in the
hole state coincides with the stationary dark soliton profile
\cite{syrwid2015}.  Note also that the analysis in
\cite{ishikawa1980} refers to an infinite noncyclic system. In this
case, classical and quantum momentums of the soliton are different.
The dispersion curves of solitons and holes are close at the weak
coupling  only in the classical definition of the soliton momentum
\cite{ishikawa1980}. If such properties hold for a cyclic system
too, then a single hole is not a soliton (despite results in
\cite{syrwid2015}), since the quantum definition of the momentum is
primary. On the whole, the connection between a hole and a soliton
is not quite clear yet
\cite{sato2012,sato2016,brand2018,syrwid2015}.

We have shown above for the weak  coupling that the hole is a
collection of identical interacting phonons with the momentum
$p=2\pi /L$. Possibly, the collection of identical phonons with
$p=4\pi /L$ (or $p=6\pi /L$, etc.) reveals also solitonic properties
at the weak coupling. Most probably, a hole has solitonic properties
only for high momenta: in this case, the hole consists of a large
number of identical phonons, and the collective effect is possible.
The solitonic properties of holes are interesting, it is worth
studying them in more details. In our opinion, it is better to use
zero boundary conditions, because $\rho(x,t)\neq const$ in this
case, and the density wave is possible.

\section{Conclusion}
We have shown that, in the case of the weak coupling, the hole with
the momentum $p=jp_{0}$ is a collection of $j$ identical interacting
phonons with the momentum $p_{0}=\pm \hbar 2\pi/L$. In this case,
the particles are elementary excitations, and the holes are
composite ones. If $j\sim N$, the hole corresponds to the condensate
of phonons. Thus, Lieb's excitations quite agree with the
Bogolyubov's and Feynman's solutions.  The traditional point of
view, according to which a holes are an independent type of
excitations, has survived for so long since the Lieb--Liniger wave
functions was not compared with the wave functions of a system of
nonpoint bosons.

At a strong coupling, the system of interacting bosons partially
acquires the fermionic properties. This is evidenced by the
solutions obtained by Girardeau, Lieb, and subsequent authors. That
is what is missing from the Bogolyubov's and Feynman's approaches.
In this case, the holes and the particles become similar to holes
and particles in a Fermi system. The structure of quasiparticles is
very unusual at $\gamma=+\infty$: It is found above that  the
low-lying particles and holes are collections of identical
phonon-holes with the momentum $p_{0}=\pm \hbar 2\pi/L$. Apparently,
only these two quasiparticles are primary indivisible excitations in
this case --- phonon-holes with the momenta $ \hbar 2\pi/L$ and $-
\hbar 2\pi/L$.

The author thanks N. Iorgov for the valuable discussion and the
anonymous referees for helpful comments. The present work is
partially supported by the National Academy of Sciences of Ukraine
(project No.~0116U003191).

\section{Appendix 1. The largest number of quasiparticles.}
Consider $N=10^{6}$ weakly interacting Bose atoms placed in a
vessel. How many quasiparticles can exist in such a system? At first
sight, the number of quasiparticles $N_{Q}$ should not be bounded
from above, since a quasiparticle is similar to a wave in the
probability field. However, it turns out that $N_{Q}\leq N$. This
can be proved by two methods.

The most simple way is to use the Lieb--Liniger equations (\ref{1}).
In the Gaudin's numbering, the creation of a quasiparticle is
equivalent to a change in some $n_{j}$ from $n_{j}=0$ to
$n_{j}=l\neq 0$. In this case, a Bogolyubov--Feynman quasiparticle
with the momentum $p=2\pi l/L$ is created. The largest number of
quasiparticles is equal to the number of  $n$'s with different $j$:
it is the number of equations in system (\ref{1}), which is equal to
the number of atoms $N$. In this case, a hole is several
Bogolyubov--Feynman quasiparticles. These properties were noted in
\cite{mtjltp2017,mt2015}.

For nonpoint bosons it is necessary to note that a wave function
(\ref{2-3}), (\ref{2-4}) describes not only a state with one
quasiparticle, but also the states with \textit{any} number of
quasiparticles. Indeed, the WF of any stationary excited state can
be written in the form $f(\textbf{r}_1,\ldots ,\textbf{r}_N)\Psi_0$.
The periodic system has a definite momentum. The general form of the
WF of a state with the total momentum $\hbar\textbf{p}$ is set by
formulae (\ref{2-3}), (\ref{2-4})  (if the number of quasiparticles
$\geq 2$, then it is necessary to make changes in (\ref{2-3}),
(\ref{2-4}) as described in Sect. 2). Therefore, the function
$f(\textbf{r}_1,\ldots ,\textbf{r}_N)$ should coincide with
$\psi_{\textbf{p}}$ (\ref{2-4}). In this case, $b_{j}$ are different
for different states. For the state with one phonon, $b_{j}\sim 1$
for all $j$. For a state with two phonons with the momenta
$\hbar\textbf{p}_{1}$ and $\hbar\textbf{p}_{2}$ we should set
$\textbf{p}=\textbf{p}_{1}+\textbf{p}_{2}$ in (\ref{2-3}),
(\ref{2-4}). In this case, $b_{j\geq 3}\sim 1$,
$b_{1}(\textbf{p}_{1},\textbf{p}_{2},N)\sim N^{-1/2}$,
$b_{2}(\textbf{q}_{1};\textbf{p}_{1},\textbf{p}_{2},N)\sim N^{-1/2}$
for $\textbf{q}_{1}\neq -\textbf{p}_{1}, -\textbf{p}_{2}$, and
$b_{2}(\textbf{q}_{1};\textbf{p}_{1},\textbf{p}_{2},N)\sim N^{1/2}$
for $\textbf{q}_{1}= -\textbf{p}_{1}, -\textbf{p}_{2}$. For a state
with three phonons we have
$\textbf{p}=\textbf{p}_{1}+\textbf{p}_{2}+\textbf{p}_{3}$. The
lowest not small coefficients $b_{j}$ should be the coefficients
$b_{3}(\textbf{q}_{1},\textbf{q}_{2};\textbf{p}_{1},\textbf{p}_{2},\textbf{p}_{3},N)$
with such $\textbf{q}_{1}$ and $\textbf{q}_{2}$, for which
$\rho_{\textbf{q}_{1}}\rho_{\textbf{q}_{2}}\rho_{-\textbf{q}_{1}-\textbf{q}_{2}-\textbf{p}}
=
\rho_{-\textbf{p}_{1}}\rho_{-\textbf{p}_{2}}\rho_{-\textbf{p}_{3}}$.
For a state with $N$ quasiparticles the relation
$\textbf{p}=\textbf{p}_{1}+\ldots+\textbf{p}_{N}$ holds, and the
coefficients $b_{j\leq N-1}$ are negligible: $b_{j\leq N-1}\sim
N^{-a_{j}}$ ($a_{j}>0$). The coefficients
$b_{N}(\textbf{q}_{1},\ldots,\textbf{q}_{N-1};\textbf{p}_{1},\ldots,\textbf{p}_{N},N)$
are not small at such $\textbf{q}_{1},\ldots,\textbf{q}_{N-1}$, for
which
$\rho_{\textbf{q}_{1}}\ldots\rho_{\textbf{q}_{N-1}}\rho_{-\textbf{q}_{1}-\ldots-\textbf{q}_{N-1}-\textbf{p}}
= \rho_{-\textbf{p}_{1}}\ldots\rho_{-\textbf{p}_{N}}$.

Formulae (\ref{2-3}), (\ref{2-4}) imply that the largest number of
quasiparticles equals $N$, since series (\ref{2-4}) contains the
terms $\rho_{-\textbf{q}_{1}}\ldots\rho_{-\textbf{q}_{j}}$ with at
most $N$ factors $\rho_{-\textbf{q}}$. The last property is caused
by that the functions $1, \rho_{-\textbf{q}_{1}}$,
$\rho_{-\textbf{q}_{1}}\rho_{-\textbf{q}_{2}}, \ldots,
\rho_{-\textbf{q}_{1}}\ldots\rho_{-\textbf{q}_{N}}$ form the
complete (nonorthogonal) collection of functions, in which any
Bose-symmetric function of the variables $\textbf{r}_{1}, \ldots,
\textbf{r}_{N}$, which can be presented as the Fourier series, can
be expanded \cite{yuv1}. Therefore, the product
$\rho_{-\textbf{q}_{1}}\ldots\rho_{-\textbf{q}_{N}}\rho_{-\textbf{q}_{N+1}}\ldots\rho_{-\textbf{q}_{N+M}}$
containing more than $N$ factors  $\rho_{-\textbf{q}}$ is reduced to
an expansion of the form $\psi_{\textbf{p}}$ (\ref{2-4}) with
$\textbf{p}=\textbf{q}_{1}+\ldots+\textbf{q}_{N+M}$. For example,
for $N=2$ we obtain
\begin{equation}
\rho_{\textbf{q}_{1}}\rho_{\textbf{q}_{2}}
\rho_{\textbf{q}_{3}}=\frac{1}{\sqrt{N}}(\rho_{\textbf{q}_{1}+\textbf{q}_{2}}
\rho_{\textbf{q}_{3}}+\rho_{\textbf{q}_{1}+\textbf{q}_{3}}
\rho_{\textbf{q}_{2}}+\rho_{\textbf{q}_{2}+\textbf{q}_{3}}
\rho_{\textbf{q}_{1}})-\frac{2}{N}\rho_{\textbf{q}_{1}+\textbf{q}_{2}+\textbf{q}_{3}}.
     \label{5-1} \end{equation}

Thus, the largest number of quasiparticles in a Bose gas, being in
some pure state $\Psi_{p}$, is equal to $N$. According to quantum
statistics, the equilibrium number of quasiparticles for the given
temperature $T>0$ is
\begin{equation}
\bar{N}_{Q}(T)=\frac{1}{Z}\int d \textbf{r}_{1}\ldots
d\textbf{r}_{N}\sum\limits_{p}e^{-E_{p}/k_{B}T}\Psi^{*}_{p}\hat{N}_{Qp}\Psi_{p}=
\frac{1}{Z}\sum\limits_{p}e^{-E_{p}/k_{B}T}N_{Qp},
      \label{srT} \end{equation}
where $Z=\sum_{p}e^{-E_{p}/k_{B}T}$,
$\{\Psi_{p}(x_{1},\ldots,x_{N})\}$ is the complete orthonormalized
set of WFs of a system with a fixed number of atoms $N$, and
$N_{Qp}$ is the number of quasiparticles in the state $\Psi_{p}$.
According to the above analysis, the value of $N_{Qp}$ is determined
by the structure of $\Psi_{p}(x_{1},\ldots,x_{N})$, and $N_{Qp}\leq
N$ for any state. Therefore, $\bar{N}_{Q}(T)< N$. At low
temperatures, the states with small $N_{Qp}$ make the main
contribution to (\ref{srT}). Therefore, the average  number of
quasiparticles is small. In this case, $\bar{N}_{Q}(T)$ increases
with $T$. It is clear that, as $T\rightarrow \infty,$ we have
$\bar{N}_{Q}(T)\rightarrow N$. Thus, in the gas at a high
temperature, the number of quasiparticles is close to the number of
atoms.  This shows how a quantum Bose system transforms into a
classical one.

\section{Appendix 2. Vakarchuk--Yukhnovskii's equations. }
The functions $a_{j}$ and $b_{j}$ from Eqs. (\ref{2-2}) and
(\ref{2-4}) satisfy the Vakarchuk--Yukhnovskii's equations
\cite{yuv2,yuv1}
\begin{equation}
E_{0}=\frac{N-1}{2}n\nu(0)- \sum\limits_{\textbf{q}\neq
0}\frac{n\nu(q)}{2}-\sum\limits_{\textbf{q}\neq
0}\frac{\hbar^{2}q^{2}}{2m} a_{2}(\textbf{q}),
     \label{9-1} \end{equation}
\begin{equation}
\frac{mn\nu(q)}{\hbar^{2}}+q^{2}a_{2}(\textbf{q})-q^{2}a^{2}_{2}(\textbf{q})-\frac{1}{N}
\sum\limits_{\textbf{q}_{1}\neq
0}a_{3}(\textbf{q},\textbf{q}_{1})\textbf{q}_{1}(\textbf{q}+\textbf{q}_{1})
- \frac{1}{2N}\sum\limits_{\textbf{q}_{1}\neq
0}a_{4}(\textbf{q},-\textbf{q}_{1},\textbf{q}_{1})q_{1}^{2}=0,
     \label{9-2} \end{equation}
\begin{eqnarray}
&&a_{3}(\textbf{q}_{1},\textbf{q}_{2})[E_{1}(\textbf{q}_{1})+E_{1}(\textbf{q}_{2})+E_{1}(\textbf{q}_{1}+\textbf{q}_{2})]
+2\textbf{q}_{1}\textbf{q}_{2}a_{2}(\textbf{q}_{1})a_{2}(\textbf{q}_{2})-\nonumber
\\ &&-2\textbf{q}_{1}(\textbf{q}_{1}+\textbf{q}_{2})a_{2}(\textbf{q}_{1})a_{2}(\textbf{q}_{1}+\textbf{q}_{2})
-2\textbf{q}_{2}(\textbf{q}_{1}+\textbf{q}_{2})a_{2}(\textbf{q}_{2})a_{2}(\textbf{q}_{1}+\textbf{q}_{2})-
\nonumber \\ &&- \frac{1}{N}\sum\limits_{\textbf{q}\neq
0}a_{5}(\textbf{q}_{1},\textbf{q}_{2},\textbf{q},-\textbf{q})q^{2}+
\frac{1}{N}\sum\limits_{\textbf{q}\neq 0}\left
[a_{4}(\textbf{q}_{1}-\textbf{q},\textbf{q}_{2},\textbf{q})
(\textbf{q}_{1}-\textbf{q})\textbf{q}+\right.\label{9-3}
\\ &&+ \left. a_{4}(\textbf{q}_{1},\textbf{q}_{2}-\textbf{q},\textbf{q})
(\textbf{q}_{2}-\textbf{q})\textbf{q}+a_{4}(\textbf{q}_{1},\textbf{q}_{2},-\textbf{q}_{1}-\textbf{q}_{2}-\textbf{q})
(-\textbf{q}_{1}-\textbf{q}_{2}-\textbf{q})\textbf{q} \right ]=0,
     \nonumber \end{eqnarray}
\begin{eqnarray}
b_{1}(\textbf{p})E(\textbf{p})=b_{1}(\textbf{p})E_{1}(\textbf{p})-
\frac{1}{N}\sum\limits_{\textbf{q}\neq
0}b_{2}(\textbf{q};\textbf{p})\frac{\hbar^{2}}{2m}(\textbf{p}+\textbf{q})\textbf{q}
-\frac{1}{N}\sum\limits_{\textbf{q}\neq
0}b_{3}(\textbf{q},-\textbf{q};\textbf{p})\frac{\hbar^{2}q^{2}}{2m},
     \label{9-4} \end{eqnarray}
\begin{eqnarray}
&&b_{2}(\textbf{q};\textbf{p})\frac{2m}{\hbar^{2}}[E_{1}(\textbf{q})+E_{1}(\textbf{p}+\textbf{q})-E(\textbf{p})]
+2b_{1}(\textbf{p})\textbf{p}\textbf{q}a_{2}(\textbf{q})-2b_{1}(\textbf{p})p^{2}a_{3}(\textbf{p},\textbf{q})-\nonumber
\\ &&-2b_{1}(\textbf{p})\textbf{p}(\textbf{p}+\textbf{q})a_{2}(\textbf{p}+\textbf{q})
- \frac{1}{N}\sum\limits_{\textbf{q}_{1}\neq
0}q_{1}^{2}b_{4}(\textbf{q}_{1},-\textbf{q}_{1},\textbf{q};\textbf{p})+\label{9-5}
\\&&+ \frac{1}{N}\sum\limits_{\textbf{q}_{1}\neq
0}\left
[b_{3}(\textbf{q}_{1},\textbf{q}-\textbf{q}_{1};\textbf{p})\textbf{q}_{1}(\textbf{q}-\textbf{q}_{1})
+b_{3}(\textbf{q}_{1},-\textbf{q}-\textbf{q}_{1}-\textbf{p};\textbf{p})\textbf{q}_{1}(-\textbf{q}_{1}-\textbf{q}-\textbf{p})
\right ]=0,
    \nonumber  \end{eqnarray}
\begin{eqnarray}
&&b_{3}(\textbf{q}_{1},\textbf{q}_{2};\textbf{p})\frac{2m}{\hbar^{2}}[E_{1}(\textbf{q}_{1})+E_{1}(\textbf{q}_{2})+E_{1}(\textbf{p}+\textbf{q}_{1}+\textbf{q}_{2})
-E(\textbf{p})]
-2b_{1}(\textbf{p})p^{2}a_{4}(\textbf{q}_{1},\textbf{q}_{2},\textbf{p})-\nonumber
\\ &-& 2b_{1}(\textbf{p})[a_{3}(\textbf{q}_{1}+\textbf{p},\textbf{q}_{2})
\textbf{p}(\textbf{q}_{1}+\textbf{p})+a_{3}(\textbf{q}_{2}+\textbf{p},\textbf{q}_{1})
\textbf{p}(\textbf{q}_{2}+\textbf{p})-a_{3}(\textbf{q}_{1},\textbf{q}_{2})
\textbf{p}(\textbf{q}_{1}+\textbf{q}_{2})]-\nonumber
\\ &-&2b_{2}(\textbf{q}_{1};\textbf{p})a_{3}(\textbf{q}_{1}+\textbf{p},\textbf{q}_{2})(\textbf{p}+\textbf{q}_{1})^{2}
-2b_{2}(\textbf{q}_{2};\textbf{p})a_{3}(\textbf{q}_{2}+\textbf{p},\textbf{q}_{1})(\textbf{p}+\textbf{q}_{2})^{2}-
\nonumber \\ &-&
2b_{2}(-\textbf{q}_{1}-\textbf{q}_{2}-\textbf{p};\textbf{p})a_{3}(\textbf{q}_{1},\textbf{q}_{2})(\textbf{q}_{1}+\textbf{q}_{2})^{2}
-\frac{1}{N}\sum\limits_{\textbf{q}_{4}\neq
0}q_{4}^{2}b_{5}(\textbf{q}_{4},-\textbf{q}_{4},\textbf{q}_{1},\textbf{q}_{2};\textbf{p})-\nonumber
\\ &
-&2a_{2}(\textbf{q}_{1})\textbf{q}_{1}[b_{2}(\textbf{q}_{2};\textbf{p})(-\textbf{q}_{2}-\textbf{p})+
b_{2}(-\textbf{q}_{1}-\textbf{q}_{2}-\textbf{p};\textbf{p})(\textbf{q}_{1}+\textbf{q}_{2})]
-\nonumber \\ &
-&2a_{2}(\textbf{q}_{2})\textbf{q}_{2}[b_{2}(\textbf{q}_{1};\textbf{p})(-\textbf{q}_{1}-\textbf{p})+
b_{2}(-\textbf{q}_{1}-\textbf{q}_{2}-\textbf{p};\textbf{p})(\textbf{q}_{1}+\textbf{q}_{2})]
-\nonumber \\ &
-&2a_{2}(\textbf{q}_{1}+\textbf{q}_{2}+\textbf{p})(\textbf{q}_{1}+\textbf{q}_{2}+\textbf{p})[b_{2}(\textbf{q}_{1};\textbf{p})(\textbf{q}_{1}+\textbf{p})+
b_{2}(\textbf{q}_{2};\textbf{p})(\textbf{q}_{2}+\textbf{p})]+\nonumber
\\ &+&
\frac{1}{N}\sum\limits_{\textbf{q}\neq
0}\textbf{q}(-\textbf{q}_{1}-\textbf{q}_{2}-\textbf{q}-\textbf{p})
b_{4}(\textbf{q}_{1},\textbf{q}_{2},-\textbf{q}_{1}-\textbf{q}_{2}-\textbf{q}-\textbf{p};\textbf{p})+
\label{9-6}
\\&+& \frac{1}{N}\sum\limits_{\textbf{q}\neq
0}\textbf{q}(\textbf{q}_{1}-\textbf{q})
b_{4}(\textbf{q}_{1}-\textbf{q},\textbf{q}_{2},-\textbf{q}_{1}-\textbf{q}_{2}-\textbf{p};\textbf{p})+\nonumber
\\ &+& \frac{1}{N}\sum\limits_{\textbf{q}\neq
0}\textbf{q}(\textbf{q}_{2}-\textbf{q})
b_{4}(\textbf{q}_{1},\textbf{q}_{2}-\textbf{q},-\textbf{q}_{1}-\textbf{q}_{2}-\textbf{p};\textbf{p})
=0.
    \nonumber  \end{eqnarray}
Here,
$E_{1}(\textbf{q})=\frac{\hbar^{2}q^{2}}{2m}(1-2a_{2}(\textbf{q}))$.
The equation for the function $a_{4}$ is given in \cite{yuv2,yuv1}.
If one of the arguments of the functions $a_{j}$ or $b_{j}$ in
(\ref{9-1})--(\ref{9-6}) is zero, then the corresponding $a_{j}$ or
$b_{j}$ should be set zero.  If we describe the state with $l\geq 2$
quasiparticles with the total momentum $\textbf{p}_{1}+\ldots +
\textbf{p}_{l}=\textbf{p}$, then it is necessary to make the
following changes in (\ref{9-4})--(\ref{9-6}):
$E({\textbf{p}})\rightarrow E({\textbf{p}_{1},\ldots,
\textbf{p}_{l}}$) and
$b_{j}(\textbf{q}_{1},\ldots,\textbf{q}_{j-1};\textbf{p})\rightarrow
b_{j}(\textbf{q}_{1},\ldots,\textbf{q}_{j-1};\textbf{p}_{1},\ldots,
\textbf{p}_{l},N)$ for all $j$.

The functions $a_{j+1}(\textbf{q}_{1},\ldots,\textbf{q}_{j})$ and
$b_{j+1}(\textbf{q}_{1},\ldots,\textbf{q}_{j};\textbf{p})$ are
invariant relative to the permutations of two any arguments
$\textbf{q}_{l}$, $\textbf{q}_{n}$. The functions
$a_{j+1}(\textbf{q}_{1},\ldots,\textbf{q}_{j})$ are also invariant
relative to the change $\textbf{q}_{l}\rightarrow
-\textbf{q}_{1}-\textbf{q}_{2}-\ldots -\textbf{q}_{j}$ for any $j$
and  $l=1,\ldots,j$. As for the functions
$b_{j+1}(\textbf{q}_{1},\ldots,\textbf{q}_{j};\textbf{p}),$ they are
invariant relative to the change $\textbf{q}_{l}\rightarrow
-\textbf{q}_{1}-\textbf{q}_{2}-\ldots -\textbf{q}_{j}-\textbf{p}$
for any $j\geq 1$, $l=1,\ldots,j$.

In works \cite{yuv2,yuv1} a one-phonon state was considered and Eqs.
(\ref{9-1})--(\ref{9-6}) were deduced for $b_{1}(\textbf{p})=1$. We
write these equations for any $b_{1}(\textbf{p})$, so that the
equations can be used to describe the states with the number of
phonons $ \geq 1$.

Equations (\ref{9-1})--(\ref{9-6}) are exact for an infinite system:
$N, V=\infty$. For a \textit{finite} system, the product
$\rho_{-\textbf{q}_{1}}\ldots\rho_{-\textbf{q}_{N}}\rho_{-\textbf{q}_{N+1}}\ldots\rho_{-\textbf{q}_{N+M}}$
($M=1,2,\ldots$) is reduced to a sum of terms, each of which
contains at most $N$ factors of the form $\rho_{-\textbf{q}}$ (see
Appendix 1). One needs to take this property  into account while
deriving the equations for $a_{j}$ and $b_{j}$, which will cause the
appearance of many additional terms in Eqs.
(\ref{9-1})--(\ref{9-6}). However, for the weak coupling, these
terms should be negligible. Apparently, they are negligible also for
a nonweak coupling. Otherwise, the transition from the solutions for
a large finite system to solutions for the infinite one would occur
by jump. However, we do not expect such a jump. One can verify that
the solutions of the Lieb--Liniger equations (\ref{1}) or (\ref{2})
do not exhibit such a jump. Those additional terms were not
considered in the literature, and we omitted them in Sections 2, 3.

     \renewcommand\refname{}


\end{document}